\documentclass[preprint,11.5pt,authoryear]{elsarticle}

\usepackage[left=2.5cm, right=2.5cm, top=2.5cm, bottom=2.5cm]{geometry}
\usepackage{amssymb}
\usepackage{amsmath}

\usepackage[ruled,vlined]{algorithm2e}
\usepackage{algorithmic}
\usepackage{lineno}
\usepackage{todonotes} 
\usepackage{graphicx}
\usepackage{float}
\usepackage{booktabs}
\usepackage{array}
\usepackage{multirow}
\usepackage{hyperref}
\usepackage{cleveref} 
\usepackage{tabularx}
\usepackage{booktabs}
\usepackage{makecell}
\usepackage{caption}
\journal{Engineering Applications of Artificial Intelligence}
\crefname{algocf}{Algorithm}{Algorithms}
\Crefname{algocf}{Algorithm}{Algorithms}
\begin{document}
\begin{frontmatter}

\title{Coupling Smoothed Particle Hydrodynamics with Multi-Agent Deep Reinforcement 
Learning for Cooperative Control of Point Absorbers} 

\author[a,b]{Yi Zhan}
\author[b]{Iván Martínez-Estévez}
\author[a]{Min Luo\corref{cor1}}
\ead{min.luo@zju.edu.cn}
\author[b]{Alejandro J.C. Crespo}
\author[c]{Abbas Khayyer}

\affiliation[a]{organization={State Key Laboratory of Ocean Sensing \& Ocean College},
            city={Zhoushan},
            postcode={316021}, 
            state={Zhejiang},
            country={China}}

\affiliation[b]{organization={Environmental Physics Laboratory, CIM-UVIGO, Universidade de Vigo},
            city={Ourense},
            postcode={34002}, 
            country={Spain}}

\affiliation[c]{organization={Department of Civil and Earth Resources Engineering, Kyoto University},
            city={Kyoto},
            postcode={615-8540}, 
            country={Japan}}

\cortext[cor1]{Corresponding author}

\begin{abstract}
Wave Energy Converters, particularly point absorbers, have emerged as one of the most promising technologies for harvesting ocean wave energy. Nevertheless, achieving high conversion efficiency remains challenging due to the inherently complex and nonlinear interactions between incident waves and device motion dynamics. This study develops an optimal adaptive damping control model for the power take-off (PTO) system by coupling Smoothed Particle Hydrodynamics (SPH) with multi-agent deep reinforcement learning. The proposed framework enables real-time communication between high-fidelity SPH simulations and intelligent control agents that learn coordinated policies to maximise energy capture. In each training episode, the SPH-based environment provides instantaneous hydrodynamic states to the agents, which output continuous damping actions and receive rewards reflecting power absorption. The Multi-Agent Soft Actor Critic algorithm is employed within a centralised-training and decentralised-execution scheme to ensure stable learning in continuous, multi-body systems. The entire platform is implemented in a unified GPU-accelerated C++ environment, allowing long-horizon training and large-scale three-dimensional simulations. The approach is validated through a series of two-dimensional and three-dimensional benchmark cases under regular and irregular wave conditions. Compared with constant PTO damping, the learned control policy increases overall energy capture by 23.8\% and 21.5\%, respectively, demonstrating the strong potential of intelligent control for improving the performance of wave energy converter arrays. The developed three-dimensional GPU-accelerated multi-agent platform in computational hydrodynamics, is extendable to other fluid–structure interaction engineering problem that require real-time, multi-body coordinated control.
\end{abstract}


\end{frontmatter}



\section{Introduction}
\label{sec:intro}
In response to growing concerns over climate change and the demand for renewable energy, wave energy has garnered increasing attention in recent years \citep{edwardsEffectDeviceGeometry2025}. Wave energy converters (WECs) are designed to harness the energy of ocean waves and convert it into usable electrical power. Various configurations of WECs have been developed and applied in practical engineering, such as oscillating water columns, overtopping devices and oscillating body systems. Among the various types of WECs, point absorbers (PAs) have shown particular promise owing to their compact design, high conversion efficiency and adaptability to a wide range of wave conditions \citep{heReviewResearchApproaches2023,koliosReliabilityAssessmentPointabsorber2018}. A typical PA consists of a floating body that oscillates with incoming waves, a mooring system that constrains its displacement and a power take-off (PTO) system that transforms mechanical motion into electrical energy.

Considerable research efforts have been devoted to enhancing the energy conversion efficiency of PAs through mathematical modelling and control optimisation. Traditional approaches include tuning the power take-off (PTO) parameters to match the natural frequency of the PA with the incident wave frequency to induce structural resonance \citep{piscopoNewOptimizationProcedure2016,haraguchiEnhancedPowerAbsorption2020}, developing model predictive control schemes for real-time optimisation based on predicted wave elevations \citep{faedoOptimalControlMPC2017,sonOptimizingOceanwaveEnergy2017}, and implementing latching control strategies \citep{babaritOptimalLatchingControl2006,henriquesLatchingControlFloating2016} that temporarily restrain the floater motion when its velocity approaches zero or when it reaches the wave trough. However, these conventional optimisation techniques typically rely on linear control theory or rule-based strategies, which are often inadequate for handling the strong nonlinear, stochastic and high-dimensional characteristics of real ocean environments. In recent years, the growing effectiveness and robustness of machine learning algorithms alongside significant improvements in computing hardware have significantly enhanced the practical capabilities of deep reinforcement learning (DRL). Consequently, DRL has evolved into a powerful framework for learning optimal control policies through interactions with complex dynamic environments and has been widely adopted in ocean engineering optimisation \citep{fanReinforcementLearningBluff2020,zhangSampleefficientPlanningcontrolFramework2025,bowesReinforcementLearningbasedRealtime2022,xieSimpleApproachWave2023}. 

Early efforts to improve the energy absorption efficiency of WECs through reinforcement learning can be traced back to the pioneering work of \cite{anderliniControlPointAbsorber2016,anderliniRealTimeReinforcementLearning2020}, who used a model-free reinforcement learning algorithm to control the resistive damping of the PTO force in a heaving point absorber. Subsequent studies have further explored RL-based optimisation of PTO systems under complex nonlinear dynamics \citep{zouOptimizationElectricityGeneration2022,wangDeepReinforcementLearningbased2024,zhanModelFreeLinearNoncausal2024}. These works commonly adopted potential flow theory to simulate wave–WEC interactions, achieving high computational efficiency. More recently, an emerging research trend involves coupling DRL algorithms with computational fluid dynamics (CFD) models to optimise WEC performance. For instance, \cite{liangEnvironmentalSensingAdaptiveOptimization2024} developed a CFD–DRL coupled framework for a single-cylinder point absorber, where a DRL agent interacted with a mesh-based CFD solver to adaptively adjust the PTO damping coefficient under irregular wave conditions. Similarly, \cite{yeAdaptiveOptimizationWave2025} proposed an SPH–DRL coupled model for an oscillating wave surge converter (OWSC), where a single DRL agent learned an adaptive PTO control policy directly from high-fidelity SPH simulations. Their results confirmed that DRL can exploit nonlinear wave–structure interactions to improve energy absorption in a single-device configuration. Although these CFD-based approaches offer improved physical realism by accounting for viscous effects and complex free-surface dynamics, they require considerably higher computational cost. As DRL training usually involves a large number of episodes to achieve convergence, most existing CFD–DRL coupling studies are confined to highly simplified two-dimensional cases to keep computational time tractable \citep{jiaStrategiesEnergyefficientFlow2025,qinLatchingControlPoint2025,yeDRLinSPHOpensourcePlatform2025a}.

Building on these developments, the majority of current research on CFD-DRL coupling models for WECs remains predominantly focused on single-agent decision-making in two-dimensional (2-D) flow domains. In contrast, practical WEC arrays operate in inherently 3-D environments and comprise multiple subsystems that require coordinated control. Each agent within the array is able to access only to local observations, such as the surrounding wave field and the motion response of its corresponding point absorber. These observations differ across agents because incident and reflected waves superimpose between adjacent PAs, producing spatially varying hydrodynamic conditions. Such scenarios require distributed and cooperative decision-making strategies. A straightforward approach might involve designing a single agent with multiple outputs to control each PA. However, this method suffers from two major drawbacks. First, as the number of controlled entities increases, the dimensionality of the input space expands rapidly, resulting in reduced training efficiency and a higher risk of convergence issues \citep{busoniuMultiagentReinforcementLearning2010}. Second, real-time communication among agents in offshore environments is often limited, making it impractical to aggregate all local observations into a centralised neural network during training or execution \citep{kouzehgarMultiAgentReinforcementLearning2020}. Another relatively straightforward approach involves directly applying a control policy trained for a single agent to all agents within the array. For example, \cite{yeAdaptiveOptimizationWave2025} attempted to extend a DRL–based PTO damping strategy-originally developed for a single WEC to a dual-WEC system. However, the overall energy capture efficiency showed limited improvement due to pronounced hydrodynamic interactions between the two devices. This suggests that coordinated control is essential in order to fully leverage the coupling effects among devices in array configurations.

Multi-agent deep reinforcement learning (MADRL) is a natural solution to address these limitations, as it distributes control across multiple agents. In MADRL frameworks, each agent learns an individual policy based solely on its own local observations while interacting with other agents and the shared environment. The centralised training with decentralised execution (CTDE) framework \citep{loweMultiAgentActorCriticMixed2017a,amatoIntroductionCentralizedTraining2024} allows agents to utilise global information during training, while maintaining decentralised and scalable control during deployment. To date, only a few studies have applied MADRL to the optimisation of WEC systems. For example, \cite{sarkarMultiAgentReinforcementLearning2022} developed a multi-agent reinforcement learning controller to modulate the reactive forces of multiple PTO units in a three-legged WEC platform, achieving coordinated control of tether tensions for enhanced power capture and reduced structural stress. However, the efficiency and feasibility of coupling MADRL with high-fidelity CFD models remain largely unexplored and warrant further investigation.

Despite these advances, an important gap remains in integrating high-fidelity CFD models with MADRL for the cooperative WEC array control. Existing SPH-DRL studies, including the work of \cite{yeAdaptiveOptimizationWave2025}, focused exclusively on OWSC single-device configurations and single-agent control. Thus, they cannot capture the multi-body hydrodynamic interactions that dominate array performance. Furthermore, current CFD-DRL coupling frameworks typically rely on Python-based communication layers, which limit computational throughput and hinder scalable multi-agent training in 3-D environments. To the best of our knowledge, no existing approach provides a fully integrated, open-source, GPU-based three-dimensional multi-agent framework specifically designed for hydrodynamic applications. This gap motivates the development of a unified, GPU-accelerated SPH–MADRL framework for coordinated PTO control in WEC arrays.

Motivated by these considerations, this study presents a fully integrated 3-D SPH–MADRL platform. This is achieved by coupling a high-fidelity SPH solver with the LibTorch \citep{paszkePyTorchImperativeStyle2019} machine-learning library, enabling real-time interaction between the physical simulation and multiple cooperative agents. The reinforcement learning components were initially implemented within an extended version of the DualSPHysics solver \citep{crespoDualSPHysicsOpensourceParallel2015,dominguezDualSPHysicsFluidDynamics2022}, referred to as DualSPHysics+ \citep{zhanDualSPHysicsEnhancedDualSPHysics2025}, which was used to generate the numerical results presented in this paper. GPU parallelisation is implemented for both the SPH simulations and neural network training, substantially improving computational efficiency and enabling large-scale 3-D simulations. The framework is applied to point absorber arrays, where agents dynamically coordinate their PTO damping actions to maximise total energy capture under nonlinear multi-body interactions. Although demonstrated here for WEC applications, the unified SPH–MADRL platform is general and can be extended to other ocean-engineering systems requiring distributed and intelligent control. All coupling models, algorithms and training tools will be released as open-source software to support reproducibility and further development.

This paper is structured as follows: \Cref{sec:numerical} introduces the governing equations and fundamental SPH formulations for simulating wave–structure interactions, along with the CTDE framework and multi-agent training algorithms. The coupling strategy between the SPH solver and the MADRL model is also described in detail. \Cref{sec:numerical_investigations} presents numerical validations of the SPH solver, as well as the model setup and results for the MADRL-based optimisation of PTO damping coefficients in a PA array. Finally, the main conclusions and future research directions are summarised in \Cref{sec:conclusions}.

\section{Numerical methods}
\label{sec:numerical}
\subsection{SPH model}
\subsubsection{Governing equations for the fluid}

The governing equations of fluid motion are the conservation laws of mass and momentum in the Navier--Stokes framework. In Lagrangian form, they are written as:
\begin{equation}
\frac{D\rho}{Dt} = - \rho \nabla \cdot \mathbf{u},
\end{equation}
\begin{equation}
\frac{D\mathbf{u}}{Dt} = -\frac{1}{\rho}\nabla p + \nu \nabla^2 \mathbf{u} + \mathbf{g},
\end{equation}
where $\rho$, $\mathbf{u}$, $p$ and $\nu$ denote the fluid density, velocity, pressure and kinematic viscosity, respectively, while $\mathbf{g}$ is the gravitational acceleration. To close the above equations, a weakly compressible equation of state \citep{antuonoFreesurfaceFlowsSolved2010,batchelorIntroductionFluidMechanics1968} is adopted:
\begin{equation}p=\frac{c_f^2\rho_0}{\beta}{\left[\left(\frac{\rho}{\rho_0}\right)^\beta-1\right]}\end{equation}
where $\rho_0$ is the reference density, $c_f$ is the numerical speed of sound and term $\beta$ denotes the polytropic constant, typically set to $\beta = 7$ in fluid simulations \citep{batchelorIntroductionFluidMechanics1968}.

The fluid domain is discretised with Lagrangian particles and solved with a widely adopted $\delta$-SPH method combining with a Riemann solver (abbreviated as $\delta$R-SPH). The discretised form of the continuity equation can be written as:

\begin{equation}
\frac{D \rho_i}{D t}
=
- \rho_i \sum_{j}
\mathbf{u}_{ij} \cdot \nabla_i W_{ij} V_j
+ \zeta_i
\end{equation}
where the subscripts $i$ and $j$ denote the target particle and the neighbouring particles located within its support radius, respectively.
The numerical speed of sound is set as $c_f = 10 u_{\max}$ to ensure a sufficiently low Mach number, and $\mathbf{u}_{ij} = \mathbf{u}_i - \mathbf{u}_j$ denotes the relative velocity between particles $i$ and $j$.
For the kernel function $W$, the fifth-order C2 Wendland kernel is adopted in this study \citep{wendlandPiecewisePolynomialPositive1995}.
A density diffusion term $\zeta_i$ proposed by \cite{fourtakasLocalUniformStencil2019} is introduced to reduce the high-frequency numerical noises in the density field.
The kernel smoothing length $h$ is defined as $h = 2 d_p$, where $d_p$ refers to the initial particle spacing.

The momentum equation is solved with a Riemann solver to control the total amount of dissipation \citep{zhangWeaklyCompressibleSPH2017,xueNovelCoupledRiemann2022}.
Meanwhile, a combination of the velocity divergence error mitigation (VEM) scheme \citep{khayyerEnhancedResolutionContinuity2023} and the hyperbolic/parabolic divergence cleaning (HPDC) scheme \citep{fourtakasDivergenceCleaningWeakly2025,triccoConstrainedHyperbolicDivergence2016} is integrated to reduce the numerical noises in the velocity divergence and thereby density and pressure field.
For a target particle $i$, the discretised momentum equation is given by:
\begin{equation}
\frac{D \mathbf{u}_i}{D t}
=
- 2 \sum_{j} m_j
\left(
\frac{p^{*}}{\rho_i \rho_j}
\right)
\nabla_i W_{ij}
+ \mathbf{g}
+ \mathbf{a}_i^{\mathrm{VEM}}
- \nabla \psi_i
\end{equation}
where $p^{*}$ denotes the interfacial pressure at the intermediate state of a particle pair, which can be obtained by solving the Riemann problem \citep{zhangWeaklyCompressibleSPH2017}.
$\mathbf{a}_i^{\mathrm{VEM}}$ denotes the VEM acceleration term, which is derived from the instantaneous velocity-divergence errors.
The auxiliary scalar field $\psi$ is introduced to formulate a damped-wave equation that diffuses and advects the divergence errors in the velocity field.
More details on the enhanced SPH model are referred to \citep{zhanDualSPHysicsEnhancedDualSPHysics2025}.

\subsubsection{Boundary condition}
The solid boundaries are modelled using multi-layer dummy particles \citep{adamiGeneralizedWallBoundary2012}, where each particle is assigned physical properties (e.g., velocity and pressure) to ensure proper interaction with the surrounding fluid particles. For a boundary particle i, the hydrodynamic pressure is estimated by Shepard interpolation from neighbouring fluid particles to ensure smooth pressure fields and consistent force transfer across interfaces, which reads:
\begin{equation}
p_i
=\frac{
\sum_{j} p_j W_{ij}
+
(\mathbf{g}-\mathbf{a}_i)\cdot
\sum_{j} \rho_j \mathbf{r}_{ij} W_{ij}
}{
\sum_{j} W_{ij}
}
\end{equation}
where $\mathbf{r}_{ij} = \mathbf{r}_i - \mathbf{r}_j$, $\mathbf{a}$ is the acceleration of boundary particles.

\subsubsection{Wave generation and absorption}
Waves are generated by imposing piston-type wave maker motions at the inflow boundary \citep{altomareLongcrestedWaveGeneration2017}. For regular waves, the displacement $x(t)$ of the wave maker is defined as:
\begin{equation}
x(t) = \frac{S_0}{2}\sin(\omega t + \theta)
\end{equation}
where $\omega$ is the angular frequency, $\theta$ is the initial phase, and $S_0$ is the stroke length determined from linear wave theory:
\begin{equation}
S_0 = \frac{H\left[\sinh(kd)\cosh(kd)+kd\right]}{2\sinh^2(kd)}
\end{equation}
with $H$, $k$ and $d$ being the wave height, wave number and water depth, respectively.

For irregular waves, the motion of the wave maker is computed by superposing multiple linear wave components whose frequencies and amplitudes follow a prescribed wave spectrum \citep{altomareLongcrestedWaveGeneration2017}. The JONSWAP spectrum \citep{hasselmannDirectionalWaveSpectra1980} is discretised into $N$ frequency components within the range $[f_{\text{start}},f_{\text{stop}}]$ where each component is assigned a random initial phase $\theta_i \in [0,2\pi]$. The angular frequency $\omega_i$ and wave height $H_i$ of the $i$-th component are calculated as:
\begin{equation}
\omega_i = 2\pi f_i,\quad H_i = 2\sqrt{2S_\eta(f_i)\Delta f}
\end{equation}
where $S_\eta(f_i)$ is the spectral density at frequency $f_i$ and $\Delta f$ is the frequency increment. The corresponding stroke length of the wave maker for each component is determined by:
\begin{equation}
S_{0,i} = \frac{H_i\left[\sinh(k_i d)\cosh(k_i d)+k_i d\right]}{2\sinh^2(k_i d)}
\end{equation}

Finally, the total displacement of the piston is obtained by linear superposition of all spectral components as:
\begin{equation}
x(t) = \sum_{i=1}^{N}\frac{S_{0,i}}{2}\sin(\omega_i t + \theta_i)
\end{equation}

To avoid wave reflections at the downstream boundary, a damping region technology is adopted \citep{altomareLongcrestedWaveGeneration2017}. Within the damping region, the particle velocity $\mathbf{u}$ is gradually reduced to zero, thereby dissipating the wave energy.

\subsubsection{Wave-driven floating object}
\label{sec:wsi}

For freely floating rigid bodies, the governing equations follow the conservation of linear and angular momentum:
\begin{equation}
\begin{gathered}
M \frac{D\mathbf{V}}{Dt}
=
\sum_{j} m_j \mathbf{f}_j
+
\mathbf{F}_t, \\
\mathbf{I} \frac{D\boldsymbol{\Omega}}{Dt}
=
\sum_{j} m_j
\left( \mathbf{r}_j - \mathbf{R}_0 \right)
\times
\mathbf{f}_j
+
\mathbf{F}_r
\end{gathered}
\end{equation}
where $M$ and $\mathbf{I}$ denote the mass and moment of inertia of the floating object, $\mathbf{V}$ and $\boldsymbol{\Omega}$ are the translational and angular velocities, $\mathbf{R}_0$ is the centre of mass, and $\mathbf{f}_j$ is the hydrodynamic force per unit mass acting on the floating particle due to fluid particle interactions. Terms $\mathbf{F}_t$ and $\mathbf{F}_r$ represent the translational and rotational damping, respectively, which account for additional dissipative effects such as viscous drag or damping force. In this study, a linear damping model is adopted to simulate the heaving motion of a floating point absorber device, where $\mathbf{F}_t$ and $\mathbf{F}_r$ are expressed as:
\begin{equation}
\mathbf{F}_t = - k_\text{p} \mathbf{v},
\qquad
\mathbf{F}_r = 0
\label{eq:kp}
\end{equation}

\subsubsection{Time-integration}

Time-integration is performed with a Symplectic position Verlet scheme \citep{dominguezDualSPHysicsFluidDynamics2022,leimkuhlerIntegrationMethodsMolecular1996} to maintain stability and energy conservation. Within each time step, the integration is split into two stages. In the first half-step, the particle positions and intermediate velocities are updated as follows:
\begin{equation}
\begin{gathered}
\mathbf{u}_i^{n+1/2}
=
\mathbf{u}_i^{n}
+
0.5 \Delta t_f \mathbf{a}_i^{n}, \\
\mathbf{r}_i^{n+1/2}
=
\mathbf{r}_i^{n}
+
0.5 \Delta t_f \mathbf{u}_i^{n}, \\
\rho_i^{n+1/2}
=
\rho_i^{n}
+
0.5 \Delta t_f \chi_i^{n}
\end{gathered}
\end{equation}
where $\chi_i$ is the material derivative of density, defined as $\chi_i = D\rho_i/Dt$. Subsequently, during the second half-step, the velocity and position at $t = n + 1$ are obtained as:
\begin{equation}
\begin{gathered}
\mathbf{u}_i^{n+1}
=
\mathbf{v}_i^{n}
+
\Delta t_f \mathbf{a}_i^{n+1/2}, \\
\mathbf{r}_i^{n+1}
=
\mathbf{r}_i^{n}
+
0.5 \Delta t_f
\left(
\mathbf{u}_i^{n+1}
+
\mathbf{u}_i^{n}
\right), \\
\rho_i^{n+1}
=
\rho_i^{n}
+
\frac{2 - \varepsilon_i^{n+1/2}}{2 + \varepsilon_i^{n+1/2}} .
\end{gathered}
\end{equation}
where
$\varepsilon_i^{n+1/2}
=
-
\left(
\frac{\chi_i^{n+1/2}}{\rho_i^{n+1/2}}
\right)
\Delta t_f$. The fluid time step $\Delta t_f$ is determined by the Courant--Friedrichs--Lewy (CFL) condition:
\begin{equation}\Delta t_f=C_\mathrm{CFL}\min_i(\sqrt{\frac{h}{\mid\mathbf{a}_i\mid}},\frac{h}{c_f})\end{equation}
where $C_\text{CFL}$  is the CFL coefficient, which is set to 0.2 in this study. 

\subsection{Deep reinforcement learning (DRL) model}
\subsubsection{Soft Actor-Critic (SAC) algorithm}

In the classical reinforcement learning framework, agents observe environmental states, execute actions, and obtain rewards through interaction with the environment. The learning objective is to obtain a policy $\pi(a|s)$ that maximises the expected cumulative discounted return:
\begin{equation}
J(\pi)
=
\mathbb{E}_{\pi}
\left[
\sum_{t=0}^{T}
\gamma^{t} r(s_t,a_t)
\right]
\end{equation}
where $s_t$ and $a_t$ denote the state and action at time step $t$, $r(s_t,a_t)$ is the reward function, $\mathbb{E}$ is the expectation operator and $\gamma \in [0,1]$ is the discount factor, which is set to $\gamma = 0.99$ in this study, a standard choice in continuous applications to ensure stable long-horizon learning.

Conventional DRL methods rely either on value-based algorithms or on the direct optimisation of parameterised policies using policy gradients. However, these methods often have limitations when applied to high-dimensional continuous control problems. These limitations include poor sample efficiency, gradient variance and unstable convergence \citep{suttonReinforcementLearningIntroduction2014}. These issues were addressed by proposing the actor-critic framework \citep{haarnojaSoftActorCriticOffPolicy2018}. Within this approach, the actor represents a parameterised policy $\pi_{\theta}(a|s)$, while the critic estimates the action-value function $Q_{\phi}(s,a)$ to provide feedback for improving the policy. The policy parameters $\theta$ are optimised following the gradient of the expected return:
\begin{equation}
\nabla_{\theta} J(\pi_{\theta})
\approx
\mathbb{E}_{s_t \sim \mathcal{D},\, a_t \sim \pi_{\theta}}
\left[
\nabla_{\theta} \log \pi_{\theta}(a_t \mid s_t)
Q_{\phi}(s_t,a_t)
\right]
\end{equation}
where $\mathcal{D}$ denotes the replay buffer, the critic parameters $\phi$ are updated by minimising the Bellman error \citep{bairdResidualAlgorithmsReinforcement1995}:
\begin{equation}
J_Q(\phi)
=
\mathbb{E}_{(s_t,a_t)\sim\mathcal{D}}
\left[
\frac{1}{2}
\left(
Q_{\phi}(s_t,a_t) - y_t
\right)^2
\right]
\end{equation}
with the temporal-difference target:
\begin{equation}
y_t
=
r(s_t,a_t)
+
\gamma
\mathbb{E}_{a_{t+1}\sim\pi}
\left[
Q_{\phi}(s_{t+1},a_{t+1})
\right]
\end{equation}

Building on this foundation, the Soft Actor-Critic (SAC) algorithm \citep{haarnojaSoftActorCriticOffPolicy2018} extends the actor-critic framework by incorporating the maximum entropy principle. Specifically, SAC introduces an entropy regularization term into the reward to encourage persistent exploration and improve policy robustness. The new objective function is defined as:
\begin{equation}
J(\pi)
=
\sum_{t=0}^{T}
\mathbb{E}_{(s_t,a_t)\sim\pi}
\left[
r(s_t,a_t)
+
\alpha \mathcal{H}\!\left(\pi(\cdot \mid s_t)\right)
\right]
\end{equation}
where $\mathcal{H}(\pi(\cdot \mid s_t)) = - \mathbb{E}_{a_t \sim \pi}\!\left[\log \pi(a_t \mid s_t)\right]$ represents the entropy of the policy, and $\alpha$ is a temperature parameter that balances reward entropy maximisations. SAC employs two independent critic networks $Q_{\phi_1}$ and $Q_{\phi_2}$, together with their corresponding target networks $Q_{\bar{\phi}_1}$ and $Q_{\bar{\phi}_2}$ to avoid value overestimation.

In practice, the policy parameters are updated by minimising the following objective:
\begin{equation}
J_{\pi}(\theta)
=
\mathbb{E}_{s_t\sim\mathcal{D},\,a_t\sim\pi_{\theta}}
\left[
\alpha \log \pi_{\theta}(a_t \mid s_t)
-
\min_{i=1,2} Q_{\phi_i}(s_t,a_t)
\right]
\end{equation}
while the critic is updated using a soft Bellman backup \citep{bairdResidualAlgorithmsReinforcement1995}:
\begin{equation}
\begin{split}
J_{Q}(\phi_i)
&=
\mathbb{E}
\left[
\left(
Q_{\phi_i}(s_t,a_t)
-
\left(
r_t
+
\gamma
\mathbb{E}_{a_{t+1}\sim\pi_{\theta}}
\left[
\min_{j=1,2} Q_{\bar{\phi}_j}(s_{t+1},a_{t+1})
\right. \right. \right. \right. \\
&\qquad \left. \left. \left. \left.
-
\alpha \log \pi_{\theta}(a_{t+1} \mid s_{t+1})
\right]
\right)
\right)^2
\right],
\quad i=1,2
\end{split}
\end{equation}

The temperature coefficient $\alpha$ is automatically adjusted during the training process to balance exploration and exploitation.

By explicitly maximising both expected return and policy entropy, SAC achieves superior stability, robustness, and sample efficiency compared with conventional actor-critic methods \citep{haarnojaSoftActorCriticAlgorithms2019}. These advantages make SAC particularly suitable for continuous control problems in ocean engineering under complex wave conditions as demonstrated in previous studies \citep{sarkarMultiAgentReinforcementLearning2022,yeDRLinSPHOpensourcePlatform2025a,cuiGatedTransformerbasedProximal2025}.

\subsubsection{Multi-agent deep reinforcement learning (MADRL)}
Compared with single-agent DRL, MADRL introduces additional challenges such as non-stationarity, scalability and the need for cooperation or competition among different agents. To address these issues, a widely adopted centralised training with decentralised execution (CTDE) framework \citep{loweMultiAgentActorCriticMixed2017a} is implemented in this study. During training, agents share information through the centralised critic networks, which evaluate the joint action-value function. At execution time, each agent acts based only on its own local observations, thereby ensuring scalability and real-time applicability.

Formally, the global environment can be modelled as a Markov decision process $\mathcal{G}$ defined by the tuple:
\begin{equation}
\mathcal{G}
=
\langle
\mathcal{S},
\{\mathcal{A}^i\}_{i=1}^{N},
\mathcal{P},
\{r^i\}_{i=1}^{N},
\gamma
\rangle
\end{equation}
where $\mathcal{S}$ is the state space, $\mathcal{A}^i$ is the action space of agent $i$, $\mathcal{P}$ is the state transition probability, $r^i$ is the individual reward function. Each agent $i$ maintains its own stochastic policy $\pi^i(a^i \mid s^i)$ while interacting within a shared fluid environment, which reads:
\begin{equation}
\pi(\mathbf{a} \mid \mathbf{s})
=
\prod_{i=1}^{N}
\pi^i(a^i \mid s^i)
\end{equation}
where $\mathbf{a} = (a^1,\ldots,a^N)$ is the joint action and $s^i$ is the local observation of agent $i$. The centralised critic approximates the global action-value function:
\begin{equation}
Q(\mathbf{s},\mathbf{a})
=
\mathbb{E}
\left[
\sum_{t=0}^{\infty}
\gamma^{t}
r(\mathbf{s}_t,\mathbf{a}_t)
\mid
\mathbf{s}_0=\mathbf{s},
\mathbf{a}_0=\mathbf{a}
\right]
\end{equation}

During execution, each agent selects actions based solely on its local policy $\pi^i(a^i|s^i)$, while coordination among agents emerges implicitly through the centralised training process. The global reward can be further decomposed into agent-specific components to encourage cooperative behaviour which reads:
\begin{equation}
r(\mathbf{s},\mathbf{a})
=
\sum_{i=1}^{N}
r^i(s,a^i,\mathbf{a}^{-i})
\end{equation}
where $a^{-i}$ represents the actions of all other agents.

Following the CTDE principle, a multi-agent soft actor-critic (MASAC) algorithm is adopted to train the actor and critic networks for each agent. Similar to SAC, MASAC incorporates entropy regularisation into the objective function to encourage exploration and improve training stability. The centralised critic minimises the soft Bellman residual:
\begin{equation}
\begin{aligned}
J_Q^i(\phi_j^i)
&=
\mathbb{E}_{(s_t,a_t,r_t^i,s_{t+1})\sim\mathcal{D}}
\left[
\left(
Q_{\phi_j^i}(s_t,a_t)
-
y_t^i
\right)^2
\right],
\quad j=1,2,\ i=1,\ldots,N,
\\
y_t^i
&=
r_t^i
+
\gamma
\mathbb{E}_{a_{t+1}\sim\pi}
\left[
\min_{j=1,2}
Q_{\bar{\phi}_j^i}(s_{t+1},a_{t+1})
-
\sum_{i=1}^{N}
\alpha_i
\log \pi_{\theta_i}
\left(
a_{t+1}^i \mid s_{t+1}^i
\right)
\right]
\end{aligned}
\label{eq:masac_critic_loss}
\end{equation}
where $\phi_j^i$ and $\bar{\phi}_j^i$ represent the parameters of the $j$-th critic network and its corresponding target network of agent $i$, respectively.

The actor network $\theta^i$ is optimised by maximising the expected return value by an entropy term:
\begin{equation}
J_{\pi}(\theta^i)
=
\mathbb{E}_{(s_t^i,a_t^i)\sim\mathcal{D}}
\left[
\alpha^i
\log \pi_{\theta_i}(a_t^i \mid s_t^i)
-
\min_{j=1,2}
Q_{\phi_j^i}(s_t,a_t)
\right]
\label{eq:policy loss}
\end{equation}

The temperature coefficient $\alpha^i$ is optimised and automatically adapted by minimizing the following objective:
\begin{equation}
J_{\alpha}^i(\alpha^i)
=
\mathbb{E}
\left[
-
\alpha_i
\left(
\log \pi_{\theta_i}(a_t^i \mid s_t^i)
+
\mathcal{H}_{\text{target}}
\right)
\right]
\label{eq:temperature loss}
\end{equation}
where the target entropy $\mathcal{H}_{\text{target}}$ is a fixed hyperparameter that determines the desired stochasticity of the policy and is typically set to the negative dimensionality of the action space. The target critic networks are updated using the soft update rule:
\begin{equation}
\bar{\phi}_j^i
\leftarrow
\tau \phi_j^i
+
(1-\tau)\bar{\phi}_j^i,
\quad
j=1,2,\ i=1,\ldots,N
\label{eq:soft update}
\end{equation}
where $\tau \in (0,1)$ is a smoothing coefficient typically set to a small value.

The hyperparameters for the MASAC algorithm used in this study are presented in \Cref{tab:masac_hyperparameters}. The training pipeline of the MASAC algorithm under the CTDE framework including the neural network update procedures is detailed in \Cref{algorithm:MASAC}. Compared with the standard single-agent SAC, MASAC explicitly addresses the non-stationarity induced by multiple learning agents. This is achieved by employing centralised critics conditioned on joint actions, while retaining decentralised policies for execution. Consequently, MASAC inherits the stability and exploration benefits of SAC, while extending its applicability to cooperative and competitive multi-agent fluid--structure interaction tasks.

\begin{table}[htbp]
\centering
\caption{Hyperparameter settings for MASAC training.}
\begin{tabular}{ll}
\hline
Hyperparameter & Value \\
\hline
Optimiser & Adam \\
Batch size & 128 \\
Replay buffer size & $10^{5}$ \\
Actor learning rate & 0.003 \\
Critic learning rate & 0.003 \\
Initial entropy coefficient $\alpha$ & 0.1 \\
Entropy coefficient learning rate $\alpha_l$ & 0.003 \\
Number of hidden layers & 3 \\
Neurons per hidden layer & $[128, 128, 64]$ \\
Discount factor $\gamma$ & 0.99 \\
Activation function & ReLU \\
Number of agents & 1, 2, 3 \\
Target entropy $\mathcal{H}_{\text{target}}$ & $-1$ \\
Target smoothing coefficient $\tau$ & 0.005 \\
\hline
\label{tab:masac_hyperparameters}
\end{tabular}
\end{table}

\begin{algorithm}[htbp]
\caption{Training pipeline of the MASAC under CTDE}
\SetAlgoLined
\LinesNumbered
\SetKwFor{For}{for}{do}{end}
\SetKwIF{If}{ElseIf}{Else}{if}{then}{else if}{else}{end}
\SetKwInput{Input}{Input}
\SetKwInput{Output}{Output}
\DontPrintSemicolon
\SetAlgoVlined
\label{algorithm:MASAC}
\Input{Policy parameters $\{\theta^i\}_{i=1}^{N}$, critic parameters $\{\phi_1^i,\phi_2^i\}_{i=1}^{N}$, target critics $\{\bar{\phi}_1^i,\bar{\phi}_2^i\}_{i=1}^{N}$, temperatures $\{\alpha^i\}_{i=1}^{N}$, replay buffer $\mathcal{D}$, hyperparameters $\gamma,\tau$}
\Output{Trained policies $\{\pi_{\theta^i}\}_{i=1}^{N}$}

\For{episode $=1,2,\ldots$}{
    Initialise environment, obtain $s_0=(s_0^1,\ldots,s_0^N)$\;
    \For{time step $t=0,1,\ldots$}{
        \tcp{Decentralised execution}
        Each agent $i$ samples $a_t^i \sim \pi_{\theta^i}(\cdot|s_t^i)$ and executes $a_t$\;
        Observe $s_{t+1}, r_t, d_t$; store $(s_t,a_t,r_t,s_{t+1},d_t)$ in $\mathcal{D}$\;
        
        \If{$|\mathcal{D}| \ge N_{\min}$}{
            Sample mini-batch $\mathcal{B} \sim \mathcal{D}$\;
            
            \tcp{Centralised critic updates}
            Sample $a_{t+1} \sim \pi_{\theta}(\cdot|s_{t+1})$ for next actions\;
            \For{each agent $i$}{
                Compute target $y_t^i$ according to \Cref{eq:masac_critic_loss}\;
                Compute critic loss $J_Q^i(\phi_j^i)$ according to \Cref{eq:masac_critic_loss}\;
                Update critic parameters $\{\phi_1^i,\phi_2^i\}_{i=1}^{N}$\;
            }
            
            \tcp{Decentralised policy and temperature updates}
            Sample $a \sim \pi_{\theta}(\cdot|s)$ for current actions\;
            \For{each agent $i$}{
                Compute policy loss $-J_\pi(\theta^i)$ according to \Cref{eq:policy loss}\;
                Compute temperature loss $J_\alpha^i(\alpha^i)$ according to \Cref{eq:temperature loss}\;
                Update policy $\theta^i$ and temperature $\alpha^i$ via gradient descent\;
            }
            
            Soft update target critic parameters $\{\bar{\phi}_1^i,\bar{\phi}_2^i\}_{i=1}^{N}$ according to \Cref{eq:soft update} for all $i,j$\;
        }
    }
}
\end{algorithm}

In summary, the DRL component of this work combines the stability and exploration capabilities of the SAC algorithm with the scalability of a multi-agent extension trained under the CTDE framework. During training, a centralised critic evaluates joint actions to mitigate non-stationarity, while each agent maintains its own decentralised policy for execution. This design enables coordinated control of multiple point absorbers while preserving real-time applicability. The next section describes how this DRL framework is integrated with the SPH solver to enable two-way interactions between fluid dynamics and intelligent agents.

\subsection{Two-way coupling of SPH and DRL}

DualSPHysics is an open-source SPH solver that has been widely employed for simulating wave–structure interactions due to its high computational efficiency on GPU architectures \citep{crespoDualSPHysicsOpensourceParallel2015,dominguezDualSPHysicsFluidDynamics2022,martinez-estevezCouplingSPHbasedSolver2023}. In the present work, the numerical results were generated using an extended version of the solver, referred to as DualSPHysics+ \citep{zhanDualSPHysicsEnhancedDualSPHysics2025}. The coupling strategy described below is solver-agnostic and fully compatible with the standard DualSPHysics architecture.
The MADRL framework is implemented using the C++ distribution of PyTorch (LibTorch) \citep{paszkePyTorchImperativeStyle2019}, which provides a mature application programming interface for neural network construction and training. LibTorch is incorporated into the SPH solver as an external dynamic library, enabling both the SPH computations and the DRL algorithms to run within a unified C++ framework. This design allows hydrodynamic data and agent actions to be exchanged directly in memory without serialisation overhead, reducing communication cost and enabling tightly integrated interactions between the high-fidelity fluid solver and the real-time DRL agents.

\begin{figure}[htbp]
\centering
\includegraphics[width=1.05\textwidth]{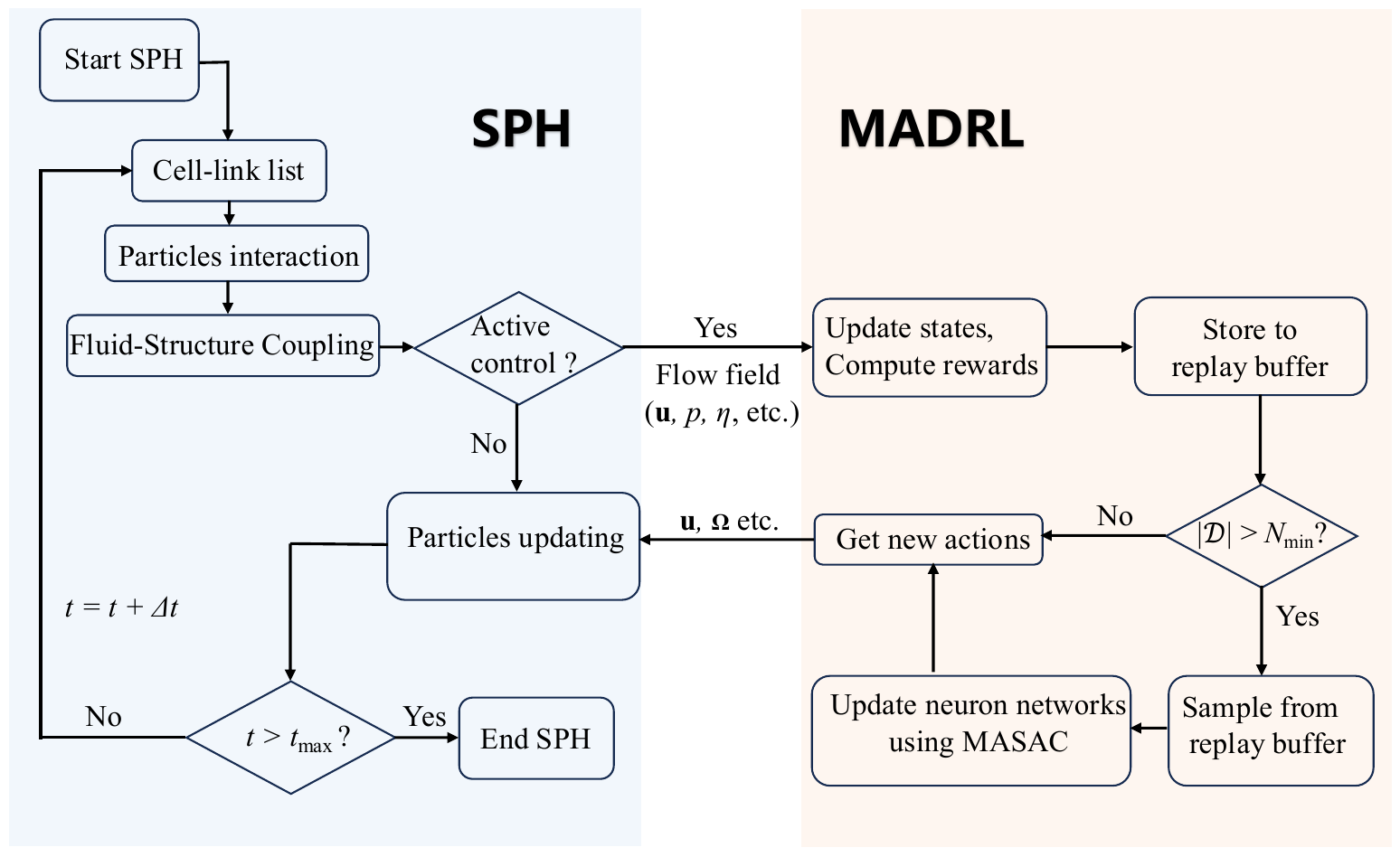} 
\caption{Coupling workflow between the SPH solver and the MADRL model.}
\label{fig:sph_drl_coupling}
\end{figure}

The two-way coupling strategy between the SPH solver and LibTorch is schematically illustrated in \Cref{fig:sph_drl_coupling} and can be divided into three main steps: 

i) the flow field information computed by the SPH solver (velocities, pressures and free-surface elevations) is transferred to the DRL neural networks as local observations;

ii) each agent generates new actions according to its neural network output, while network updates are performed conditionally depending on the filling ratio of the replay buffer as described in \Cref{algorithm:MASAC};

iii) these actions are fed back into the SPH solver to actively control the structure and then regulate the ongoing simulation.

\section{Numerical investigations}
\label{sec:numerical_investigations}
\subsection{Validation of the SPH solver}
\begin{figure}[htbp]
\centering
\includegraphics[width=1.0\textwidth]{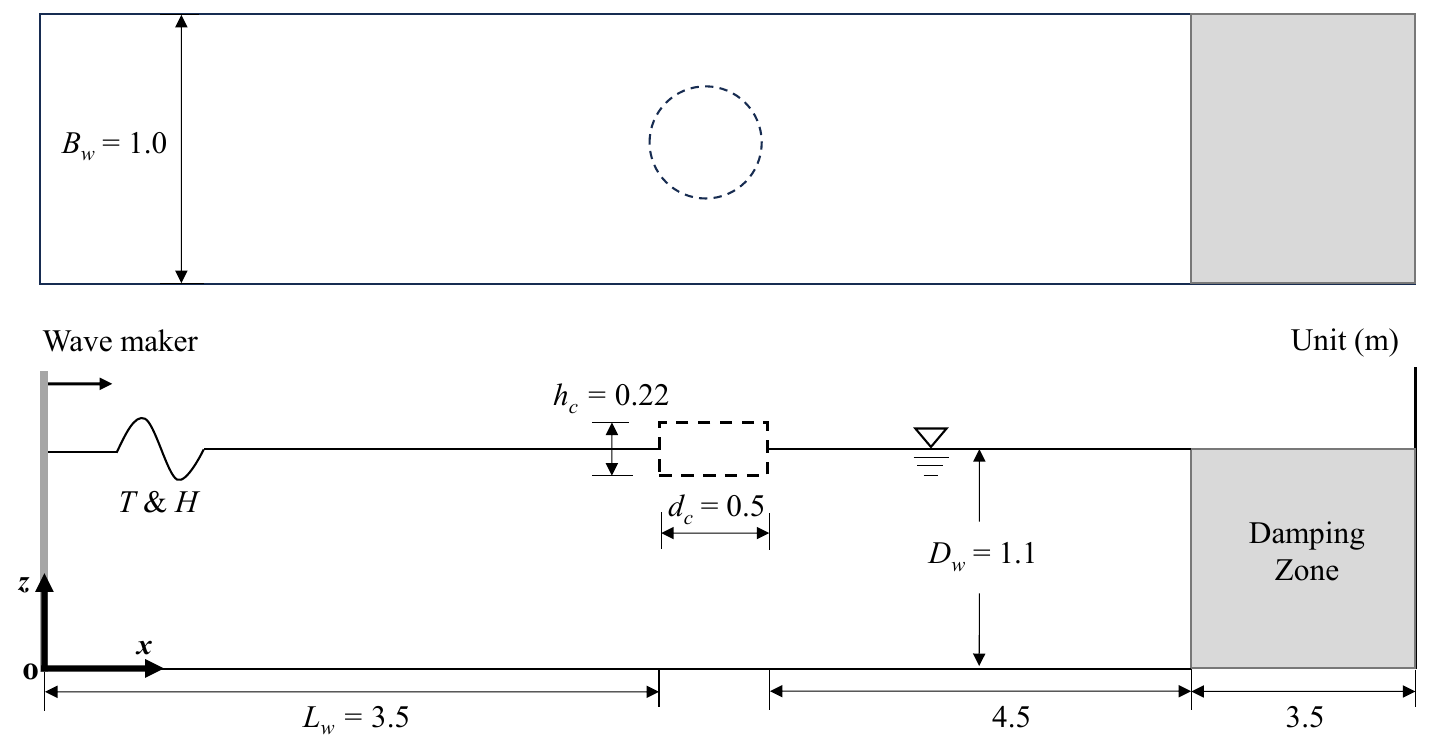}  
\caption{Numerical wave tank configuration and body dimensions.}
\label{fig:sk1}
\end{figure}

In this section, the accuracy of the SPH solver in simulating wave generation, propagation and interaction with a floating body is validated against the experimental results of \cite{zangHydrodynamicResponsesEfficiency2018}. \Cref{fig:sk1} shows the schematic view of the case setup, where the wave tank has a length of $L_w$ = 12 m, a width of $B_w$ = 1.0 m and a water depth of $D_w$ = 1.1 m. The cylinder type point absorber with a diameter $d_c$ = 0.5 m, a height $h_c$ = 0.22 m and a material density   is installed along the central midline of the tank, at a distance of $L_w$ = 3.5 m from the wave maker. 

\subsubsection{Wave generation and propagation}
\label{sec:wave_generation_propagation}

To validate the accuracy of the solver in reproducing both regular and irregular waves, preliminary simulations are conducted in an empty wave tank without any floating bodies. Waves are generated by a wave maker at the left boundary and propagate toward the right. A wave gauge is positioned at the centre of the tank to record free-surface elevations. For the regular wave case, the wave height and period are set to $H = 0.16$~m and $T = 1.5$~s, respectively. For the irregular wave case, the JONSWAP spectrum \citep{hasselmannDirectionalWaveSpectra1980} is employed with a significant wave height of $H_s = 0.16$~m and a peak period of $T_s = 1.5$~s. The irregular wave train is composed of 50 superimposed wave components.

\begin{figure}[htbp]
\centering
\includegraphics[width=1.0\textwidth]{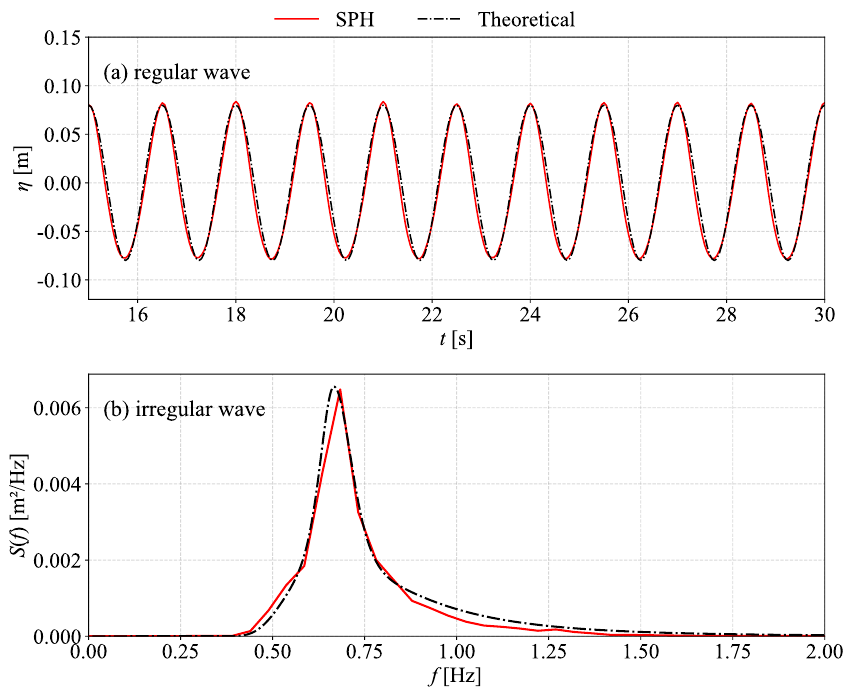}  
\caption{Comparison between SPH simulations and theoretical solutions for wave generation: (a) free-surface elevation for the regular waves and (b) irregular wave spectrum density.}
\label{fig:validation_wave_generation}
\end{figure}

\Cref{fig:validation_wave_generation} compares the SPH simulations with the corresponding theoretical solutions for wave-generation validation. The wave elevation time series of regular waves is presented in \Cref{fig:validation_wave_generation}(a). As can be seen, the numerical results show a good agreement with the theoretical solution, with maximum relative errors of 3.17\% at the crest and 2.77\% at the trough. \Cref{fig:validation_wave_generation}(b) shows the SPH-simulated spectrum density together with the target JONSWAP spectrum density over a 200~s wave generation--propagation interval. The significant wave height ($H_s$) and peak period ($T_s$) obtained from SPH simulations are 0.151~m and 1.463~s, respectively, showing relative errors of 5.63\% and 2.48\% compared to the theoretical values of 0.16~m and 1.5~s. Overall, these results demonstrate that the SPH model can accurately reproduce both regular and irregular wave conditions considered in this study.

\subsubsection{Wave--floating body interaction}
\label{sec:wave_floating_body_interaction}

\begin{figure}[htbp]
\centering
\includegraphics[width=1.05\textwidth]{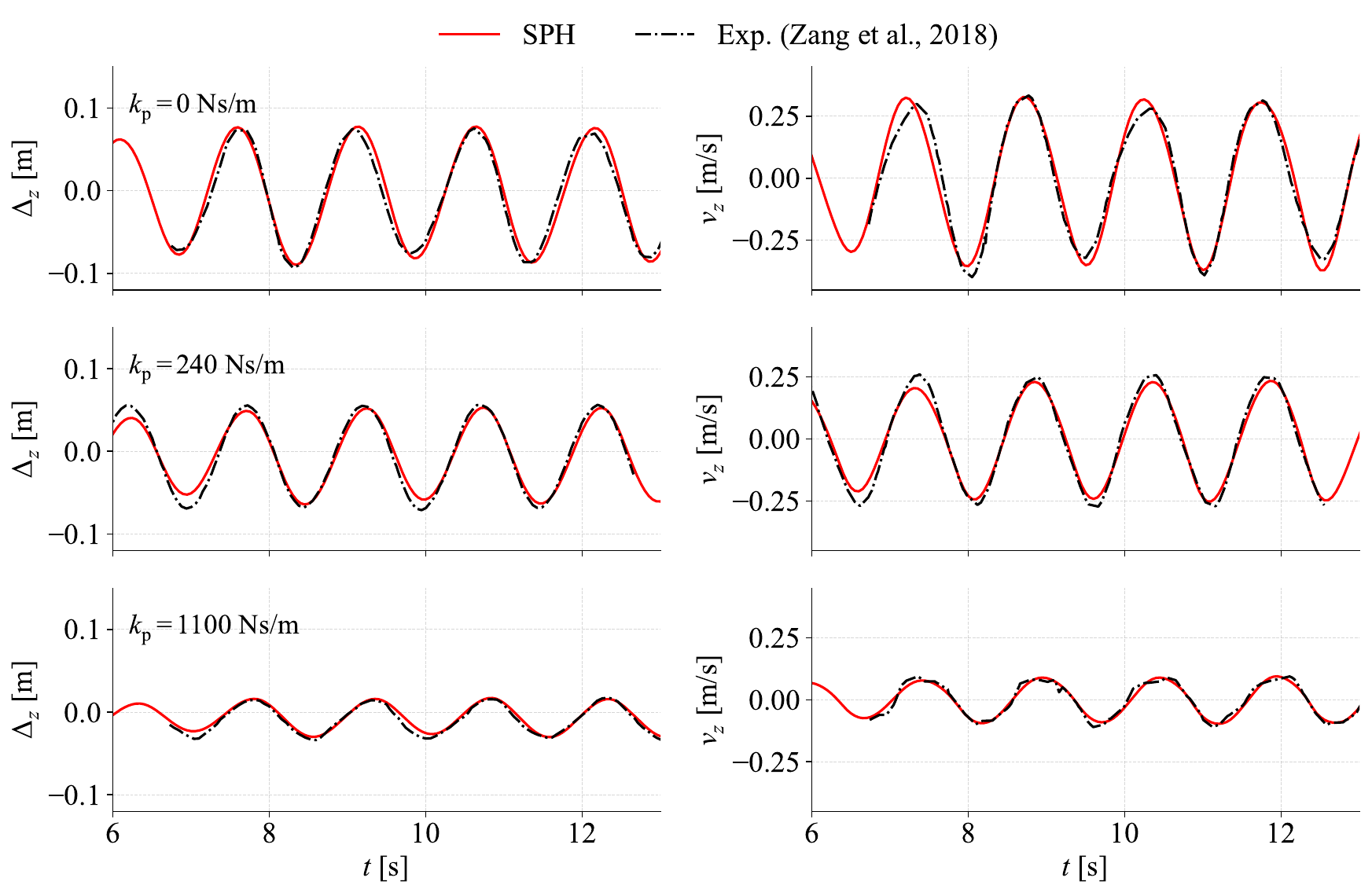}  
\caption{Time histories of the simulated heave displacement and velocity of the point absorber together with the experimental data \citep{zangHydrodynamicResponsesEfficiency2018}.}
\label{fig:validation_fsi}
\end{figure}

Subsequently, wave--floating WEC interactions are simulated. The regular wave condition previously validated ($H = 0.16$~m, $T = 1.5$~s) is employed here, consistent with the experimental setup of \cite{zangHydrodynamicResponsesEfficiency2018}. The PTO force applied in the SPH model follows the linear damper formulation used in the validation study of \cite{ropero-giraldaEfficiencySurvivabilityAnalysis2020}, defined in \Cref{eq:kp}, where $k_\text{p}$ is the damping coefficient and $v_z$ is the heave velocity. Therefore, simulations are also conducted for three constant PTO damping coefficients: $k_\text{p} = 0$, 240, and 1100~Ns/m. \Cref{fig:validation_fsi} compares the time histories of the heave displacement $\Delta z$ and velocity $v_z$ obtained from SPH simulations with experimental data. The results show good agreement with the experimental data for all three damping cases, indicating that the SPH solver can reliably reproduce the wave--floating body response under the tested conditions.

\subsection{Adaptive optimisation of PTO damping in PA arrays}
\label{sec:adaptive_optimisation_pto_pa_arrays}

\begin{figure}[htbp]
\centering
\includegraphics[width=1.05\textwidth]{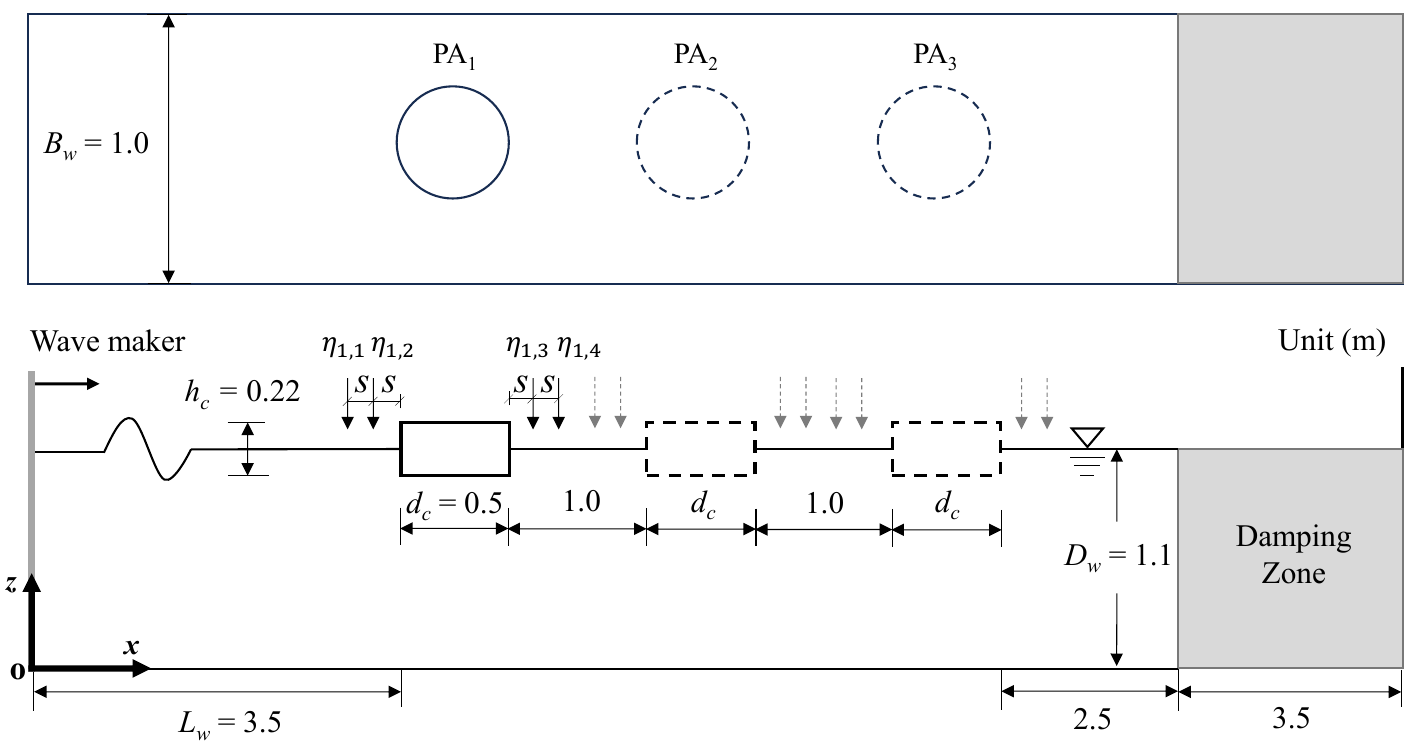}  
\caption{Numerical wave tank setup and point absorber array layout.}
\label{fig:sk2}
\end{figure}

In this section, the damping coefficient of the PTO system (i.e., $k_\text{p}$ in \Cref{eq:kp}) is dynamically optimised using the SPH-MADRL model to improve the energy harvesting efficiency of both single-device and multi-device arrays. \Cref{fig:sk2} shows the schematic view of the case setup. The dimensions of both the wave flume and the point absorbers are identical to those in \Cref{fig:sk1}. Multiple PAs are arranged along the central axis of the flume with a spacing of 1.0~m between adjacent devices and are labelled sequentially as PA1, PA2, and PA3 from left to right. For each PA, two wave gauges are arranged upstream and two downstream to record the local wave elevations. The spacing between adjacent gauges is $S = 0.1$~m. A 2-D regular-wave case is first simulated to validate the robustness and effectiveness of the coupled model, while also providing a simplified setting to investigate how the DRL controller enhances the energy absorption efficiency of the WEC. Subsequently, a 3-D irregular-wave case, which is closer to realistic sea conditions, is simulated to demonstrate the model’s computational efficiency and practical effectiveness in fully 3-D applications.

In this case, simulations are conducted with one-to-three point absorbers to evaluate the capability of the proposed model in handling both single-agent and multi-agent control scenarios. Each agent $i$ receives a local observation $s_i$ that includes wave elevations measured at four points around the device $(\eta_{i,1}, \eta_{i,2}, \eta_{i,3}, \eta_{i,4})$ and their temporal derivatives $(D\eta_{i,1}/Dt$, $D\eta_{i,2}/Dt$, $D\eta_{i,3}/Dt$, $D\eta_{i,4}/Dt)$, along with the heave velocity from the previous time step $v^{n-1}_{z,i}$, heave displacement from the initial position $\Delta z_i$, and the heave acceleration $a_{z,i}$. The action corresponds to the adjustment of the PTO damping coefficient $k_{\text{p},i}$ for each device. Specifically, the actor network outputs a normalised scalar action $o_i \in [-1,1]$, which is mapped to the physical damping value by:
\begin{equation}
k_{\text{p},i} = k_{\text{base}} + o_i \Delta k_{\max}, \quad \Delta k_{\max} = 0.9 k_{\text{base}}
\label{eq:krange}
\end{equation}
where $k_{\text{base}}$ is the baseline coefficient and $\Delta k_{\max}$ is the maximum allowable deviation. To encourage exploration of higher-reward control strategies, $\Delta k_{\max}$ is set to a relatively wide range, i.e., $\Delta k_{\max} = 0.9 k_{\text{base}}$. Between two consecutive training intervals, the value of $k_{\text{p},i}$ is linearly interpolated at each SPH time step to ensure smooth temporal evolution. The energy output $P$ of each device at time step $n$ can be computed as:
\begin{equation}
P_i^{n} = k_{\text{p},i}^{n} \left( \frac{v_{z,i}^{n} + v_{z,i}^{n-1}}{2} \right)^2
\end{equation}

It is evident that higher values of $P_i^{n}$ result in greater instantaneous power output. Accordingly, the reward function is first designed to provide positive reinforcement based on the energy output of each device. Furthermore, to capture cooperative interactions among multiple point absorbers within the array, the energy outputs of neighbouring devices are also incorporated. Based on these backgrounds, the reward function for agent $i$ is defined as:
\begin{equation}
r_i = (1 - \gamma_p) P_i + \gamma_p \frac{1}{N} \sum_{j=1}^{N} P_j
\end{equation}
where $N$ denotes the number of agents, the weighting parameter $\gamma_p$ is set to $\gamma_p = 0.7$ in this section.

Note that for a given point absorber geometry and specific wave conditions, there typically exists an optimal fixed value of the PTO damping coefficient, denoted as $k_0$, which maximises the energy output. In this study, $k_{\text{base}}$ in \Cref{eq:krange} is set equal to $k_0$, aiming to enhance energy extraction beyond this baseline through adaptive control. To determine $k_0$ in the 2-D case, a numerical parametric study is conducted by varying the PTO damping coefficient $k_\text{p}$ within the range $[200, 1800]$~Ns/m, which is consistent in magnitude with the values reported in an existing study \citep{ropero-giraldaEfficiencySurvivabilityAnalysis2020}. The average absorbed power $\bar{P}$ is computed as:
\begin{equation}
\bar{P} = \frac{1}{T} \int_{t_0}^{t_0+T} P(t)\,dt, \quad P(t) = k_\text{p} v_z^2(t)
\end{equation}

\begin{figure}[htbp]
\centering
\includegraphics[width=1.0\textwidth]{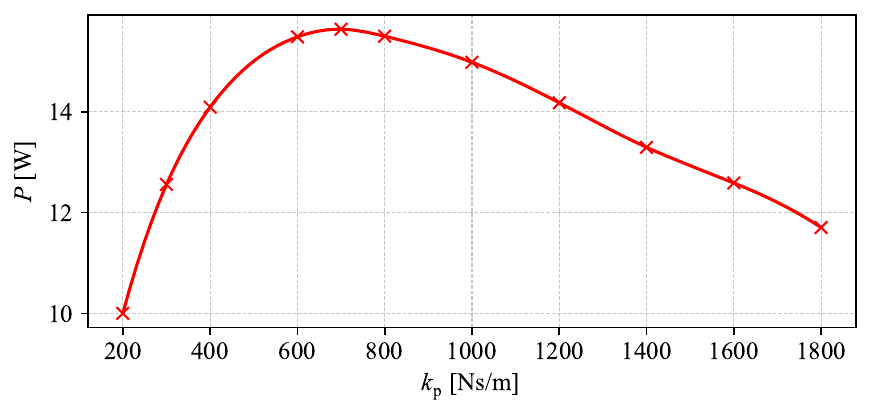}  
\caption{Variation of average power output with respect to PTO coefficient for a PA under 2-D regular wave conditions.}
\label{fig:kpto_vs_power}
\end{figure}

\Cref{fig:kpto_vs_power} presents the variation of $\bar{P}$ with respect to $k_\text{p}$. The results show that $\bar{P}$ increases initially with $k_\text{p}$ and then gradually decreases, with the maximum energy output achieved at $k_\text{p} = 700$~Ns/m. Therefore, the baseline value is set to $k_{\text{base}} = k_0 = 700$~Ns/m in the 2-D case. For the 3-D case considered in this section, the optimal value $k_{\text{base}} = k_0 = 400$~Ns/m is adopted based on the existing numerical studies \citep{ropero-giraldaEfficiencySurvivabilityAnalysis2020,manawaduSPHbasedNumericalModelling2024}.

In this study, the training process is organised into episodes. Each episode has a duration $T_e$, within which a sufficient number of wave periods must be included to ensure that the full interaction between the PA and the incident waves is captured, thereby enabling the collection of an adequate number of training samples. At the end of each episode, the simulation environment is reset, all SPH particle information is restored to their configuration at time instant $t_e$ and the next episode begins. In this work, $t_e$ is defined as a moment at which the incoming wave has fully interacted with all point absorbers. The total number of episodes $N_e$ is selected to provide sufficient training duration for policy convergence while keeping the computational cost tractable. Accordingly, the episode parameter settings for the above cases are summarised in \Cref{tab:episode}. To evaluate the robustness and reproducibility of the training process, each configuration is trained independently three times, following a standard practice in the existing literature \citep{xieSloshingSuppressionActive2021,maniaSimpleRandomSearch2018,yeAdaptiveOptimizationWave2025}. After training is completed, an additional sequence of 10 episodes is simulated to evaluate the performance of the learned policy.

\begin{table}[htbp]
\centering
\caption{Episode configuration parameters for the DRL training}
\renewcommand{\arraystretch}{1.2}
\begin{tabular}{lccc}
\toprule
 & \makecell{Episode \\ duration \\ $T_e$ (s)} & \makecell{Initial time \\ of each episode \\ $t_e$ (s)} & \makecell{Total number \\ of episodes \\ $N_e$} \\
\midrule
2-D regular wave case & 10 & 10 & 100 \\
3-D irregular wave case & 20 & 10 & 50 \\
\bottomrule
\label{tab:episode}
\end{tabular}
\end{table}

\subsubsection{Two-dimensional regular waves}

\begin{figure}[htbp]
\centering
\includegraphics[width=1.0\textwidth]{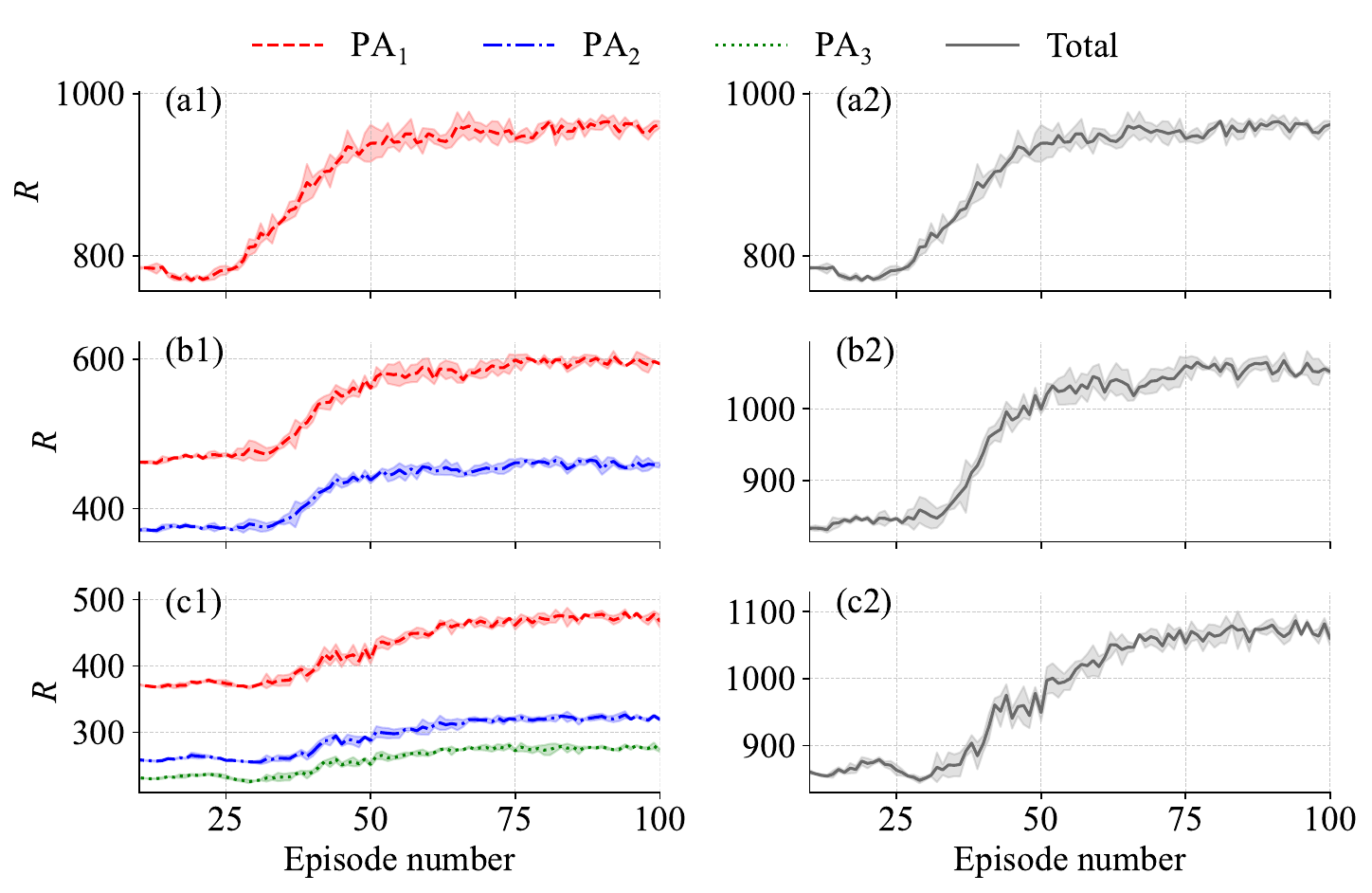}  
\caption{Time history of the training rewards for the one-, two-, and three-PAs cases.}
\label{fig:rew_wec_rewards}
\end{figure}

Firstly, a 2-D point absorber array under regular wave conditions is simulated to validate the efficiency of the proposed model. The training reward curves with shaded standard deviations for the one-, two-, and three-PAs cases are shown in \Cref{fig:rew_wec_rewards}. Note that because each agent selects random actions during the first 10 episodes, the associated rewards are not informative. Therefore, only results obtained after episode 10 are reported. As shown in the figures, all three cases exhibit a rapid reward increase around episode 20, indicating the agents are learning policies that can improve the cumulative reward. After about 60 episodes, the rewards in all cases stabilise with only slight fluctuations, demonstrating the convergence of the learning process. In the multi-agent cases, the rewards of $\textrm{PA}_1$, $\textrm{PA}_2$, and $\textrm{PA}_3$ decrease sequentially, reflecting the progressive reduction in available wave energy for downstream absorbers. Nevertheless, as shown in rows (b) and (c), the final reward values remain substantially higher than their initial levels, indicating that each agent successfully learns an effective control strategy despite differences in local wave conditions.

\begin{figure}[htbp]
\centering
\includegraphics[width=1.0\textwidth]{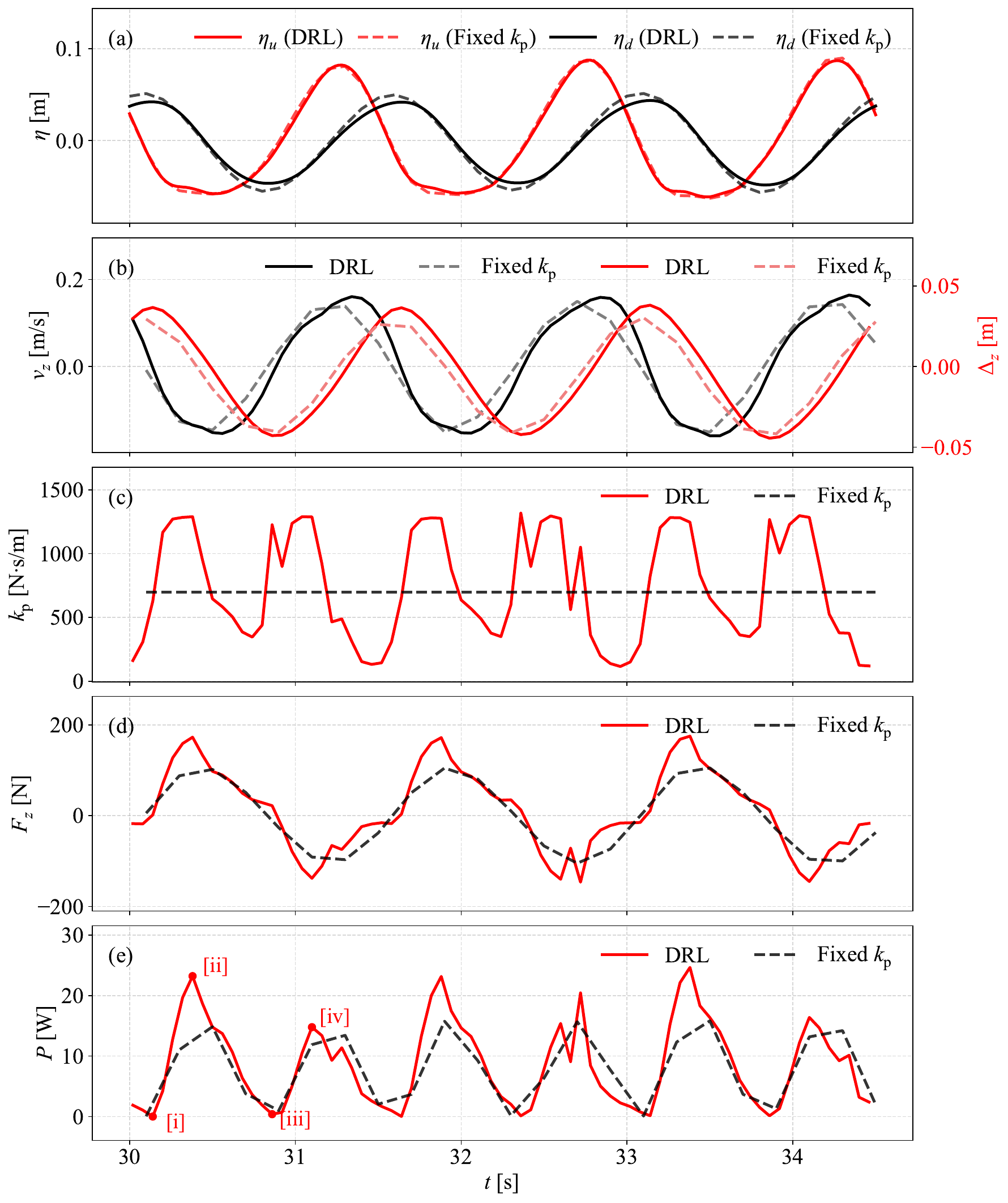}  
\caption{Time histories of the wave elevations at 0.05 m upstream ($\eta_u$) and downstream ($\eta_d$) of the PA together with the heave displacement ($\Delta_z$), velocity ($v_z$), PTO force ($F_z$) and instantaneous power output ($P$) at three typical wave periods.}
\label{fig:re_1wec_five}
\end{figure}

To evaluate the energy harvesting performance of the trained agents, the results obtained using the trained PA are compared with those obtained using a fixed PTO coefficient, i.e., $k_\text{p} = k_0 = 700$~Ns/m. The time histories of the wave elevations at 0.05~m upstream $(\eta_u)$ and downstream $(\eta_d)$ of the PA together with the heave displacement $(\Delta z)$, velocity $(v_z)$, PTO force $(F_z)$, and instantaneous power output $(P)$ at three typical wave periods are presented in \Cref{fig:re_1wec_five}. As can be seen, under nearly identical incident wave conditions, the downstream wave elevations of the DRL-trained point absorber are lower than those obtained with a constant PTO damping coefficient. The trained PA also exhibits larger heave velocities and displacements at both the wave crest and trough, indicating that the DRL control strategy effectively regulates the motion to enhance wave energy absorption. The optimised PTO damping coefficient $k_\text{p}$ shows a periodic variation, characterised by two peaks and two troughs within each wave period. Specifically, $k_\text{p}$ increases during the approach of the wave crest, leading to a higher PTO force and thus greater energy output; it then decreases during the transition from crest to trough, allowing the PA to accelerate more freely. As the wave trough approaches, the heave velocity of the PA again reaches a maximum value, accompanied by another peak in $k_\text{p}$, which further enhances the instantaneous power capture. Overall, these results demonstrate that the DRL algorithm used here effectively learns an adaptive damping strategy at different wave phases, thereby enhancing the PA's dynamic response and overall energy harvesting efficiency.

\begin{figure}[htbp]
\centering
\includegraphics[width=1.0\textwidth]{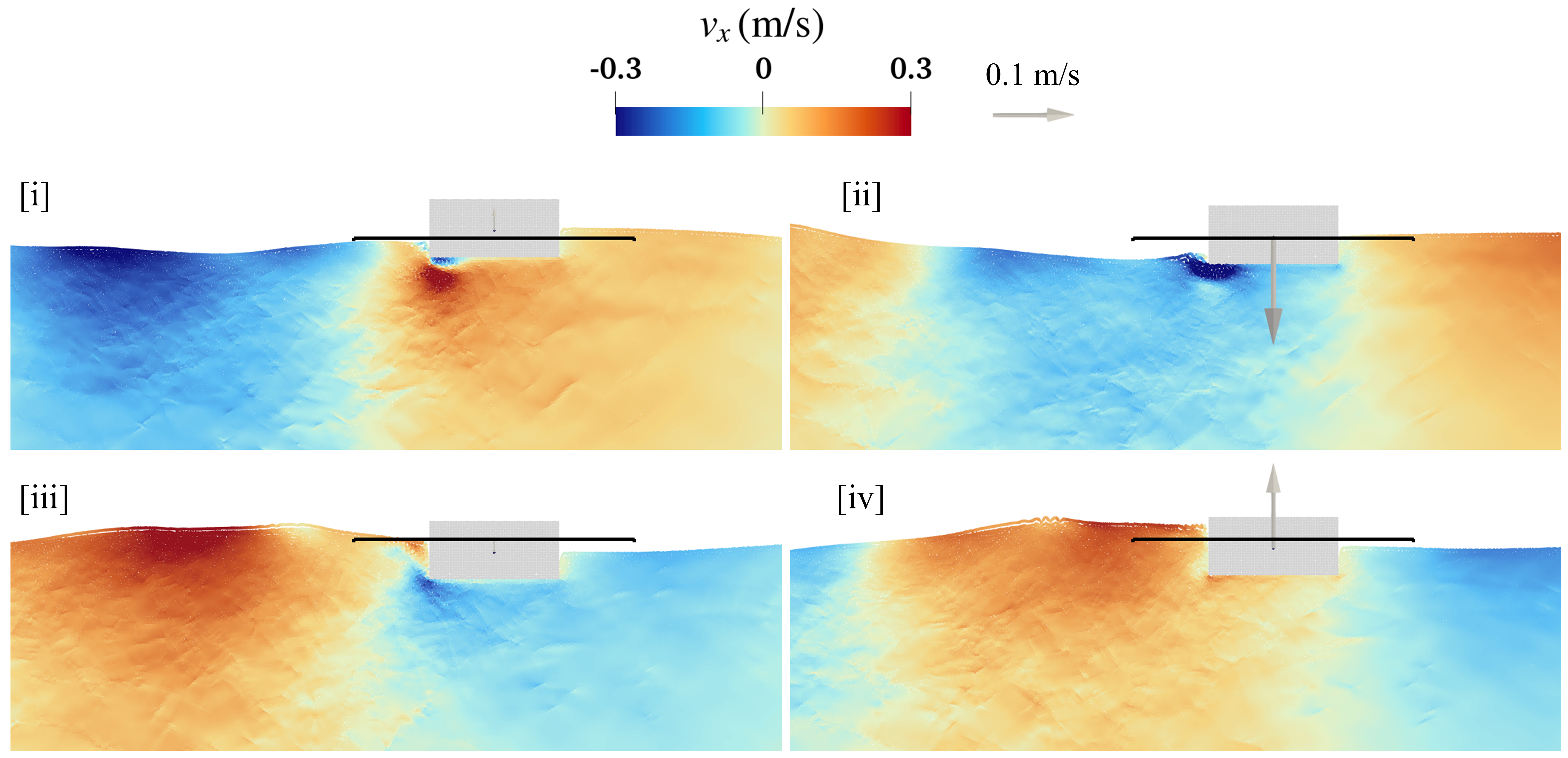}  
\caption{Snapshots of the fluid velocity field and the PA heave velocity at four typical time instants.}
\label{fig:1WEC}
\end{figure}

To further investigate the mechanism by how the DRL-trained policy enhances energy capture efficiency in this 2-D cases, \Cref{fig:1WEC} presents the velocity fields and the corresponding motion velocities of the floating body at four representative time instants (the black line indicates the still water level), which correspond to the time instants [i]--[iv] marked in \Cref{fig:re_1wec_five}(e). It can be observed that when the wave crest approaches, both the heave velocity and displacement of the PA increase rapidly, and the blocking effect of the PA induces a significant water elevation on the upstream side. After the crest passes, the PA begins to submerge and accelerates downward under gravity to a velocity even higher than that during the crest approach. Overall, the PA's motion response follows the variation trend of the wave elevation but exhibits a small phase lag. The DRL-trained control policy adapts to this behaviour by increasing the damping coefficient in anticipation of the approaching crest or trough, at which time the PA's velocity increases under wave excitation, leading to enhanced hydrodynamic damping force and improved energy-capture efficiency.

\begin{figure}[htbp]
\centering
\includegraphics[width=1.0\textwidth]{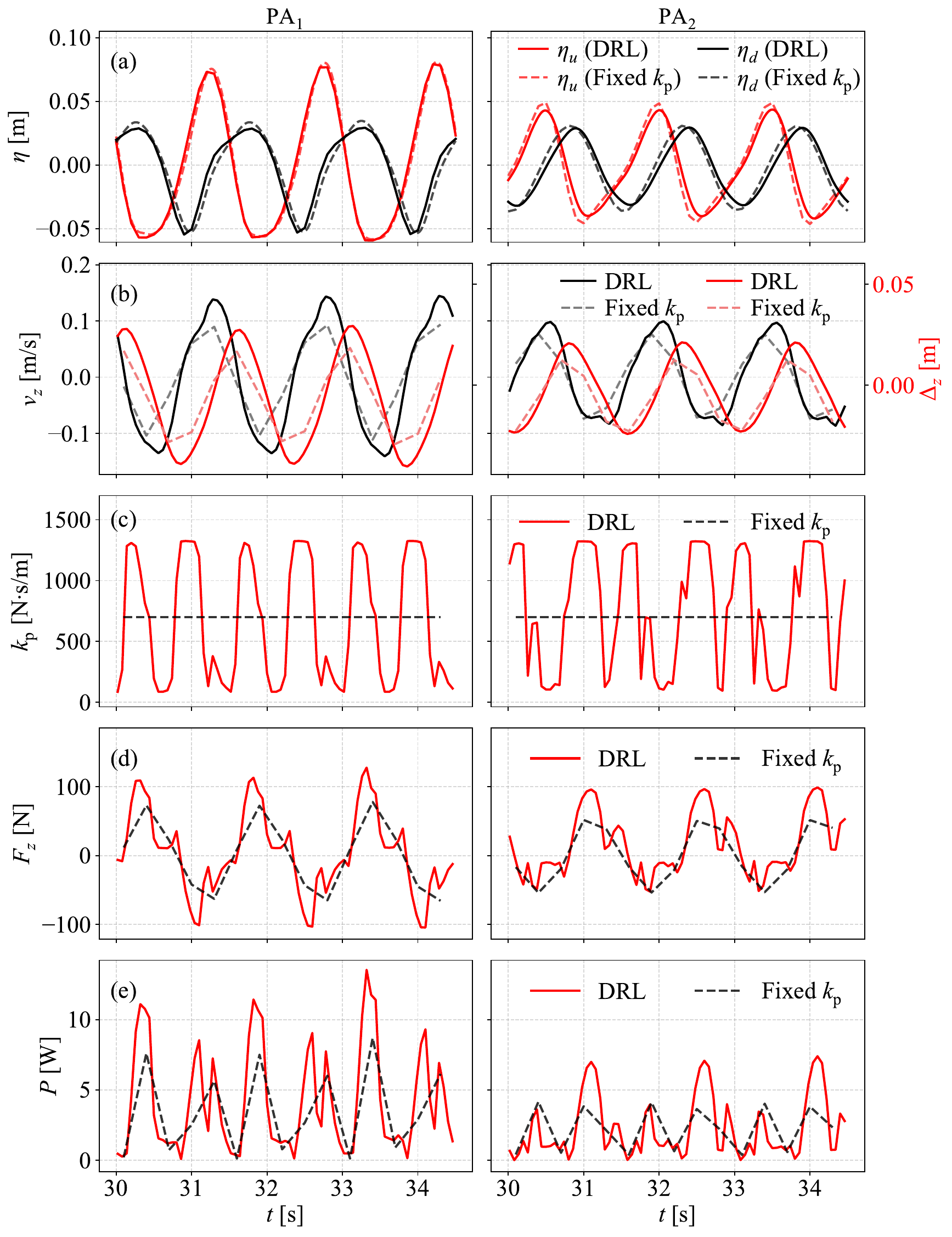}  
\caption{Time histories of the wave elevations at 0.05 m upstream ($\eta_u$) and downstream ($\eta_d$) of two PAs together with the heave displacement ($\Delta_z$), velocity ($v_z$), PTO force ($F_z$) and instantaneous power output ($P$) at three typical wave periods.}
\label{fig:re_2wecs_five}
\end{figure}

\begin{figure}[htbp]
\centering
\includegraphics[width=1.0\textwidth]{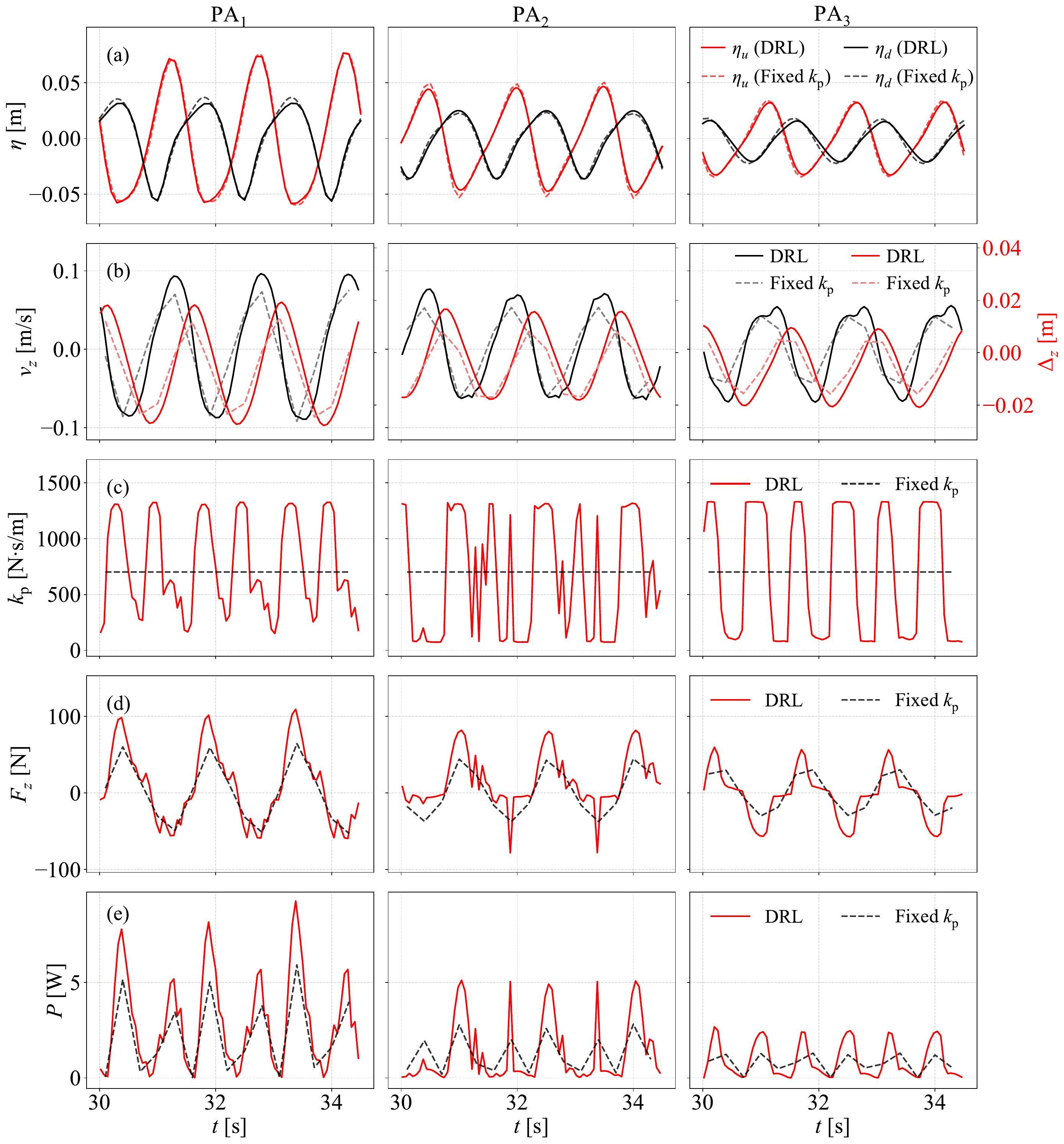}  
\caption{Time histories of the wave elevations at 0.05 m upstream ($\eta_u$) and downstream ($\eta_d$) of three PAs together with the heave displacement ($\Delta_z$), velocity ($v_z$), PTO force ($F_z$) and instantaneous power output ($P$) at three typical wave periods.}
\label{fig:re_3wecs_five}
\end{figure}

\Cref{fig:re_2wecs_five} and \Cref{fig:re_3wecs_five} show the time histories of the wave elevations at 0.05~m upstream $(\eta_u)$ and downstream $(\eta_d)$ of the PA together with the heave displacement $(\Delta z)$, velocity $(v_z)$, PTO force $(F_z)$, and instantaneous power output $(P)$ in two and three PAs cases, respectively. It can be observed that the free-surface elevation, motion response, PTO force, and power output all exhibit a gradual attenuation from upstream to downstream (i.e., from PA1 to PA3), primarily because the upstream PAs extract part of the incident wave energy, reducing the energy available to downstream devices. Similar to the single PA case, the DRL-trained PAs in both cases exhibit lower downstream free-surface elevations compared to the constant PTO baseline under nearly identical incident wave conditions. The heave velocities and displacements of the trained PAs are also larger at both the wave crest and trough, indicating that this DRL control strategy effectively regulates the motion to enhance wave energy absorption. The optimised PTO damping coefficients $k_{\text{p},i}$ for each PA show periodic variations, characterised by two peaks and two troughs within each wave period. Specifically, $k_{\text{p},i}$ increases during the approach of the wave crest, leading to higher PTO forces and thus greater energy outputs; it then decreases during the transition from crest to trough, allowing the PAs to accelerate more freely. As the wave trough approaches, the heave velocities of the PAs again reach maximum values, accompanied by another peak in $k_{\text{p},i}$, which further enhances instantaneous power capture. In summary, these results demonstrate that this DRL algorithm effectively learns adaptive damping strategies for each PA at different wave phases, thereby enhancing their dynamic responses and overall energy harvesting efficiencies in multi-device configurations.

\begin{figure}[htbp]
\centering
\includegraphics[width=1.0\textwidth]{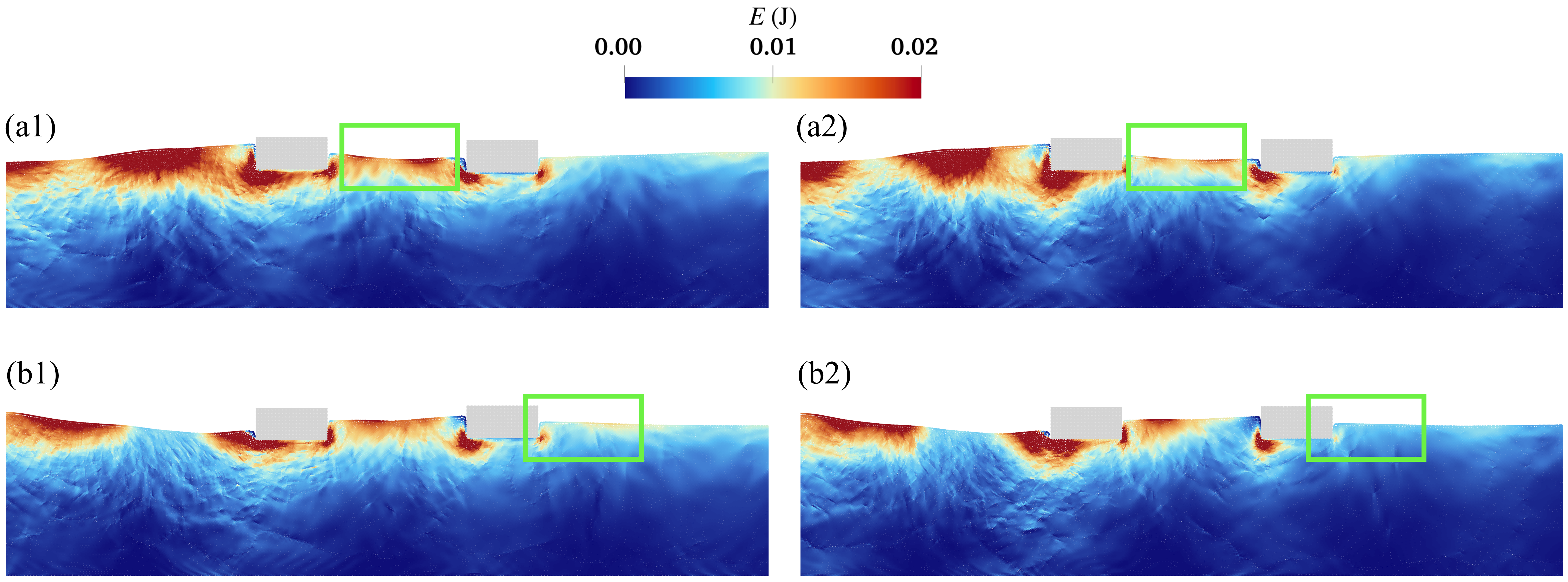}  
\caption{Snapshots of the wave-induced kinetic energy distribution around the two PAs at two representative time instants: (a) $t$ = 32.8 s and (b) $t$ = 33.5 s. Left: constant $k_\text{pto}$ baseline; right: DRL-trained case.}
\label{fig:2WEC}
\end{figure}

\begin{figure}[htbp]
\centering
\includegraphics[width=1.0\textwidth]{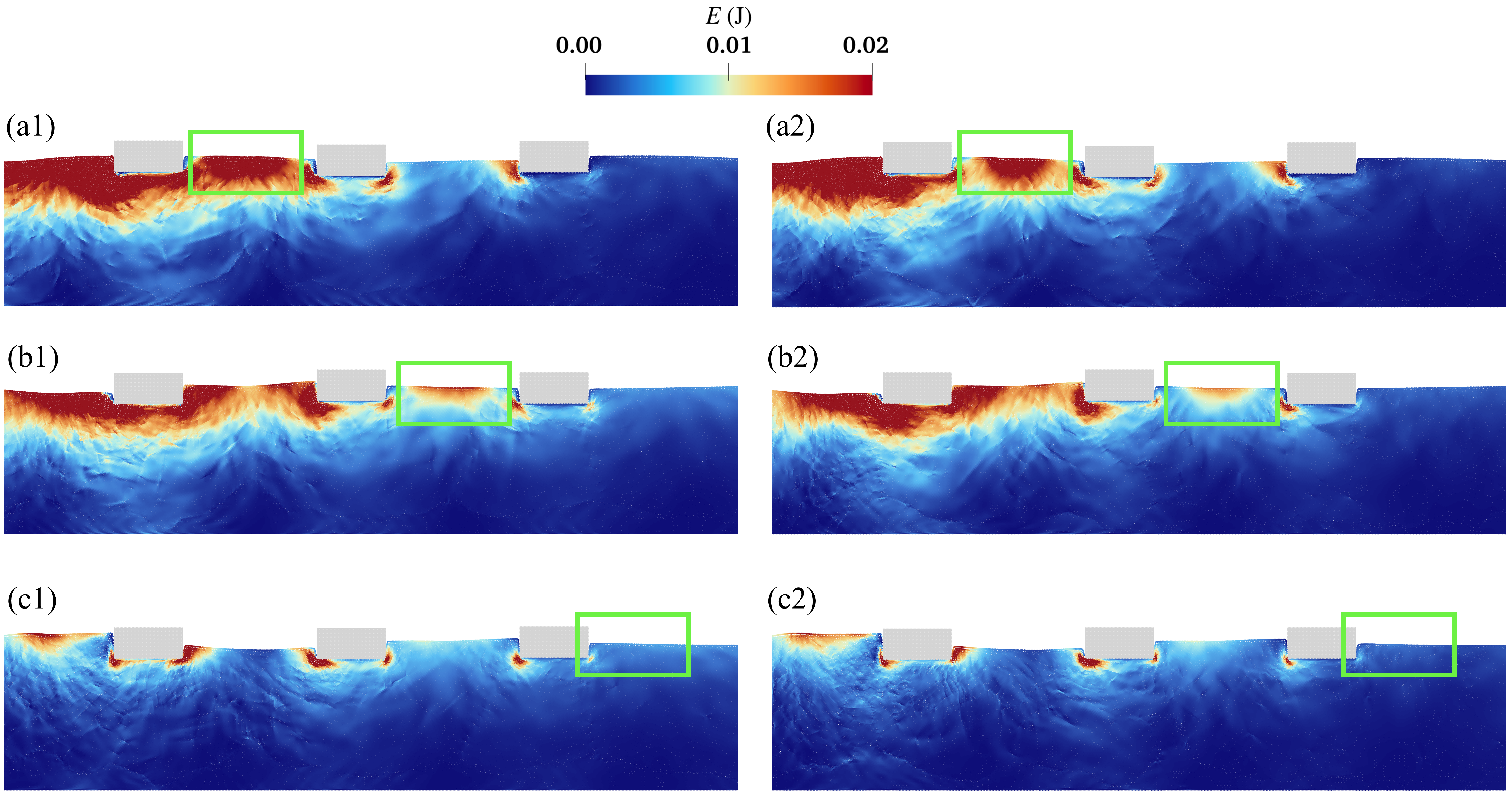}  
\caption{Snapshots of the wave-induced kinetic energy distribution around the three PAs at three representative time instants: (a) $t$ = 31.5 s, (b) $t$ = 32.1 s and (c) $t$ = 32.7 s. Left: constant $k_\text{pto}$ baseline; right: DRL-trained case.}
\label{fig:3WEC}
\end{figure}

\Cref{fig:2WEC} shows the kinetic energy distribution around the PAs at two representative time instants (i.e., $t = 32.8$~s and $33.5$~s), corresponding to the arrival of the incident wave crest at the first and second WECs, respectively. It can be observed that when the wave crest approaches the first PA, the fluid in front of the device exhibits high kinetic energy. After the crest passes PA$_1$, most of the wave energy is absorbed, leading to a significant reduction in the kinetic energy of the fluid in its wake. However, a comparison between subfigures (a1) and (a2) reveals that the fluid downstream of the DRL-trained PA exhibits lower kinetic energy compared to the constant-PTO case, indicating more effective wave energy absorption. Similarly, the comparison between subfigures (b1) and (b2) demonstrates that the DRL-trained PA$_2$ also achieves higher energy-capture efficiency. In the three-PA configuration, \Cref{fig:3WEC} depicts the kinetic energy distribution at three representative time instants ($t = 31.5$, $32.1$ and $32.7$~s), corresponding to the passage of the wave crest across PA$_1$, PA$_2$, and PA$_3$, respectively. The results show that most of the wave energy is absorbed by PA$_1$ and subsequently decreases from PA$_1$ to PA$_3$. Nevertheless, the regions highlighted by green boxes indicate that the DRL-trained PAs consistently exhibit lower downstream kinetic energy, confirming their enhanced energy absorption capability and validating the effectiveness of the DRL-trained control strategy.

\Cref{tab:re_energy_comparison} quantitatively compares the average power output of the active-controlled and constant PTO cases under different numbers of point absorbers. In all constant cases, the PTO coefficient is fixed at $k_0 = 700$~Ns/m as discussed in \Cref{sec:adaptive_optimisation_pto_pa_arrays}. For the single point absorber case, the agent achieves a 9.2\% increase in power output compared with the constant PTO baseline. In the multi-agent cases, the average power output of individual absorbers generally decreases from the first to the last unit due to the limited wave energy available in the two-dimensional domain. Energy extracted by upstream devices reduces the incident wave amplitude available to downstream units, while the superposition of radiated and incident waves further complicates the local flow field. Nevertheless, the trained agents adapt to these interactions and achieve higher overall energy extraction compared with the constant PTO baseline. In the two-absorber case, both units exhibit notable improvements of 21.7\% and 14.4\%, respectively, leading to a total energy enhancement of 18.9\%. In the three-absorber case, all agents benefit from the active control strategy, with the first, second, and third devices achieving 33.7\%, 8.9\%, and 19.1\% improvements, respectively. This corresponds to an overall energy gain of 23.8\% relative to the constant PTO configuration.

\begin{table}[htbp]
\centering
\caption{Comparison of accumulated absorbed energy under different numbers of point absorbers.}
\label{tab:re_energy_comparison}
{\renewcommand{\arraystretch}{1.2}  
\begin{tabular}{llcccc}
\toprule
Scenario & Agent & $E_{\text{DRL}}$ (J) & $E_0$ (J) & $\Delta E$ (J) & Improvement (\%) \\
\midrule
\multirow{2}{*}{One PA}
& PA$_1$ & 39.16 & 35.87 & 3.29 & 9.2 \\
& Total & 39.16 & 35.87 & 3.29 & 9.2 \\
\midrule
\multirow{3}{*}{Two PAs}
& PA$_1$ & 19.25 & 15.82 & 3.43 & 21.7 \\
& PA$_2$ & 11.17 & 9.76 & 1.41 & 14.4 \\
& Total & 30.42 & 25.58 & 4.84 & 18.9 \\
\midrule
\multirow{4}{*}{Three PAs}
& PA$_1$ & 13.30 & 9.94 & 3.36 & 33.7 \\
& PA$_2$ & 5.99 & 5.50 & 0.49 & 8.9 \\
& PA$_3$ & 4.18 & 3.51 & 0.67 & 19.1 \\
& Total & 23.47 & 18.95 & 4.52 & 23.8 \\
\bottomrule
\end{tabular}}
\end{table}

\subsubsection{Three-dimensional irregular waves}

\begin{figure}[htbp]
\centering
\includegraphics[width=1.0\textwidth]{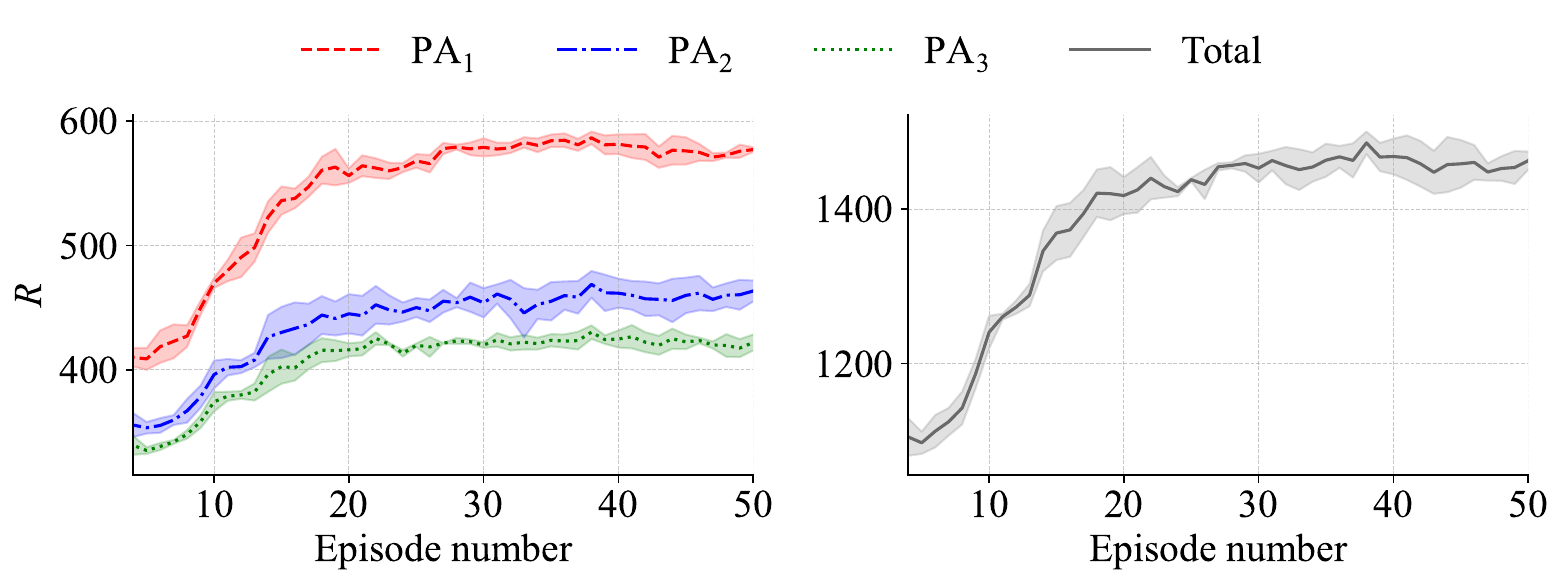}  
\caption{Time histories of the rewards for three PAs under 3-D irregular wave conditions.}
\label{fig:irre_wec_rewards}
\end{figure}

In this section, a 3-D case of a three-point absorber array under irregular wave conditions is simulated to further evaluate the efficiency of the proposed model in a more realistic environment. Firstly, the convergence and stability of the training process are validated. \Cref{fig:irre_wec_rewards} shows the time histories of the individual agents' rewards as well as the total reward. All reward curves increase rapidly during the first 20 training episodes and reach stable levels around episode 30, with obvious improvements compared to their initial values, indicating that the training process has converged. The value of the rewards decreases from the first PA to the last one because the wave energy is progressively absorbed along the array.

\begin{figure}[htbp]
\centering
\includegraphics[width=1.0\textwidth]{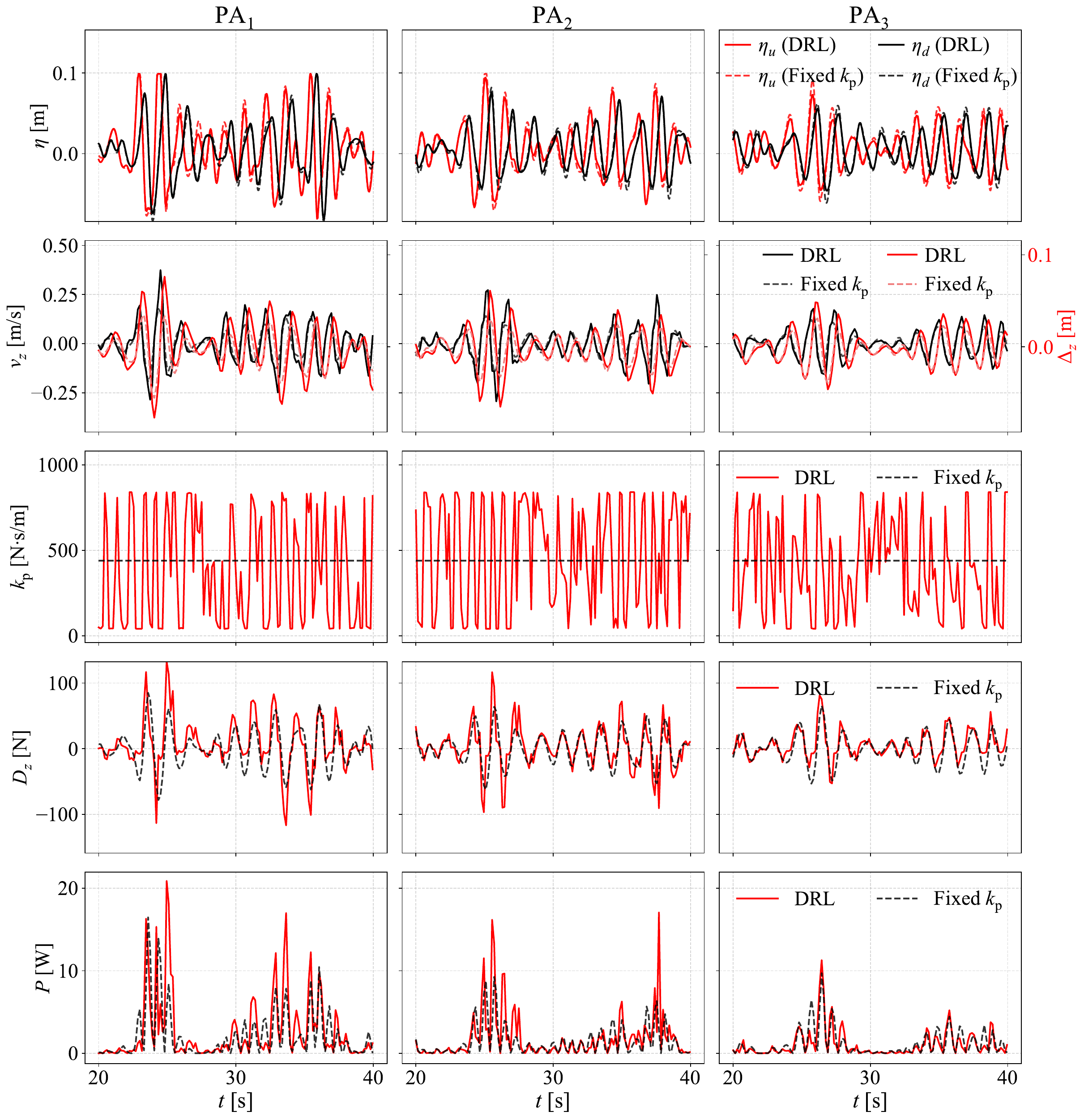}  
\caption{Time histories of the wave elevations at 0.05 m upstream ($\eta_u$) and downstream ($\eta_d$) of three PAs together with the heave displacement ($\Delta_z$), velocity ($v_z$), PTO force ($F_z$) and instantaneous power output ($P$) under 3-D irregular waves over a typical 20 s duration.}
\label{fig:irre_3wecs_five}
\end{figure}

\Cref{fig:irre_3wecs_five} shows the time histories of the free-surface elevations measured at 0.05~m upstream $(\eta_u)$ and downstream $(\eta_d)$ of each PA, together with the heave displacement $(\Delta z)$, velocity $(v_z)$, PTO force $(F_z)$, and instantaneous power output $(P_{\text{out}})$. Similar to the case under regular waves, the DRL-trained strategy enhances energy absorption primarily by adjusting the damping coefficient $k_\text{p}$, which increases the PA motion response and PTO force near each wave crest and trough. The main difference under irregular wave conditions is that the variation of $k_\text{p}$ becomes less periodic, exhibiting higher-frequency fluctuations due to the nonuniform wave periods. Correspondingly, the power output curve shows pronounced peaks associated with higher incident waves, where the capture efficiency improvement achieved by DRL control becomes more evident.

\begin{figure}[htbp]
\centering
\includegraphics[width=1.0\textwidth]{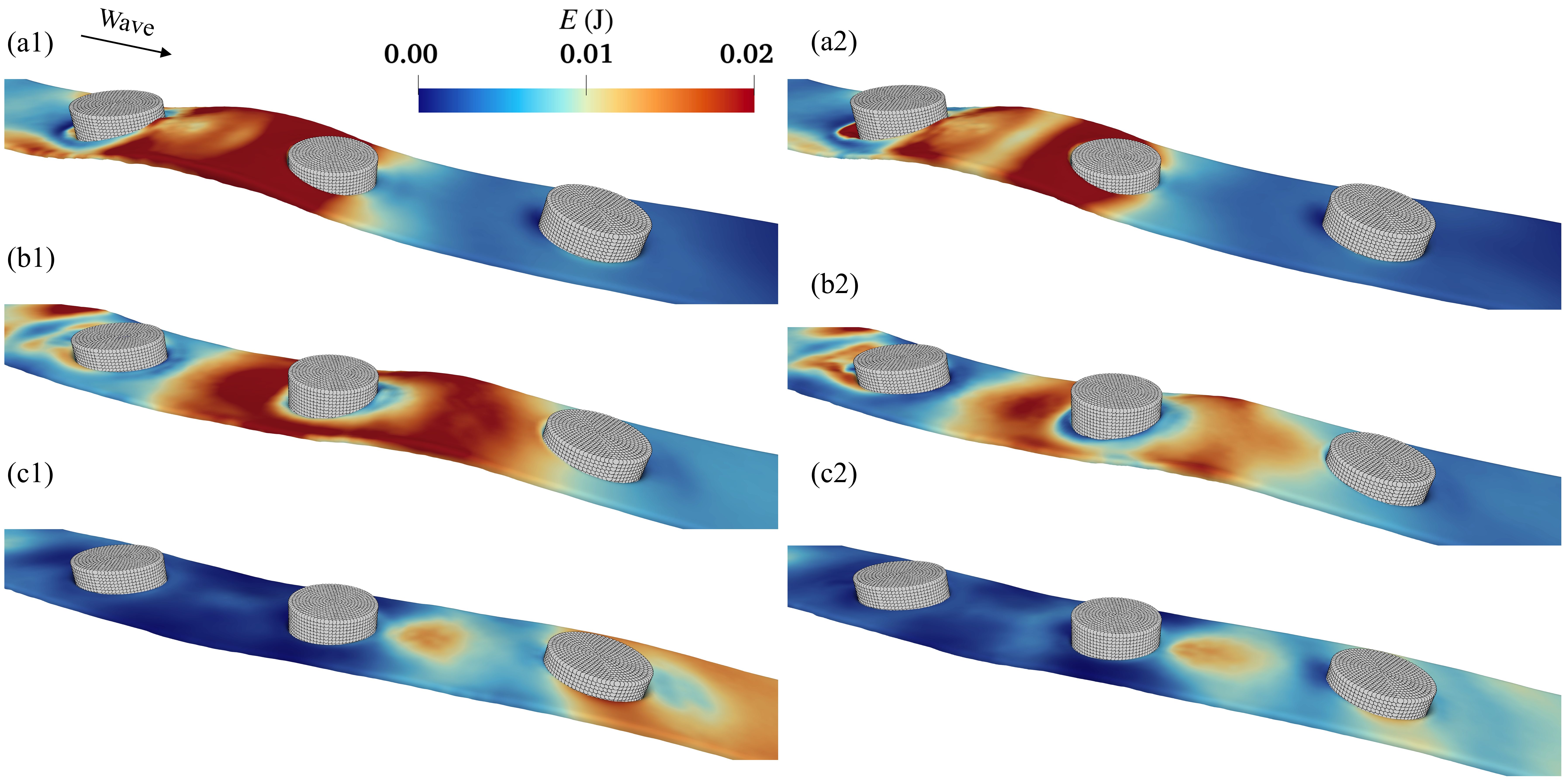}  
\caption{Snapshots of the kinetic energy distribution around the three PAs at three representative time instants under 3-D irregular wave conditions: (a) $t$ = 24.9 s, (b) $t$ = 25.6 s, and (c) $t$ = 27.3 s. Left: constant $k_\text{pto}$ baseline; right: DRL-trained case.}
\label{fig:irre3WEC_energycontour}
\end{figure}

\Cref{fig:irre3WEC_energycontour} shows the wave energy field at three representative time instants (i.e., $t = 24.9$, $25.6$ and $27.3$~s), corresponding to the arrival of the incident wave crest at the first, second and third WECs, respectively. As illustrated in subfigures (a1) and (a2), when the wave crest passes PA$_1$, part of the wave energy is absorbed by the device. The wave energy behind the DRL-trained PA$_1$ is markedly lower than that in the untrained case, indicating a higher energy-capture efficiency. When the crest reaches PA$_2$, the energy carried by the fluid in front of the untrained PA$_2$ remains higher than that of the trained one due to the upstream energy absorption by PA$_1$. Nevertheless, the comparison between subfigures (b1) and (b2) reveals that the DRL-trained PA$_2$ also exhibits superior energy-harvesting capability. After passing PA$_1$ and PA$_2$, most of the wave energy has been absorbed; however, the untrained PA$_3$ still fails to effectively capture the remaining energy. In contrast, the DRL-trained PA$_3$ leaves almost no residual wave energy in its wake, demonstrating that the coordinated actions of the trained multi-agent system significantly enhance the overall energy-absorption efficiency of the array.

\begin{figure}[htbp]
\centering
\includegraphics[width=1.0\textwidth]{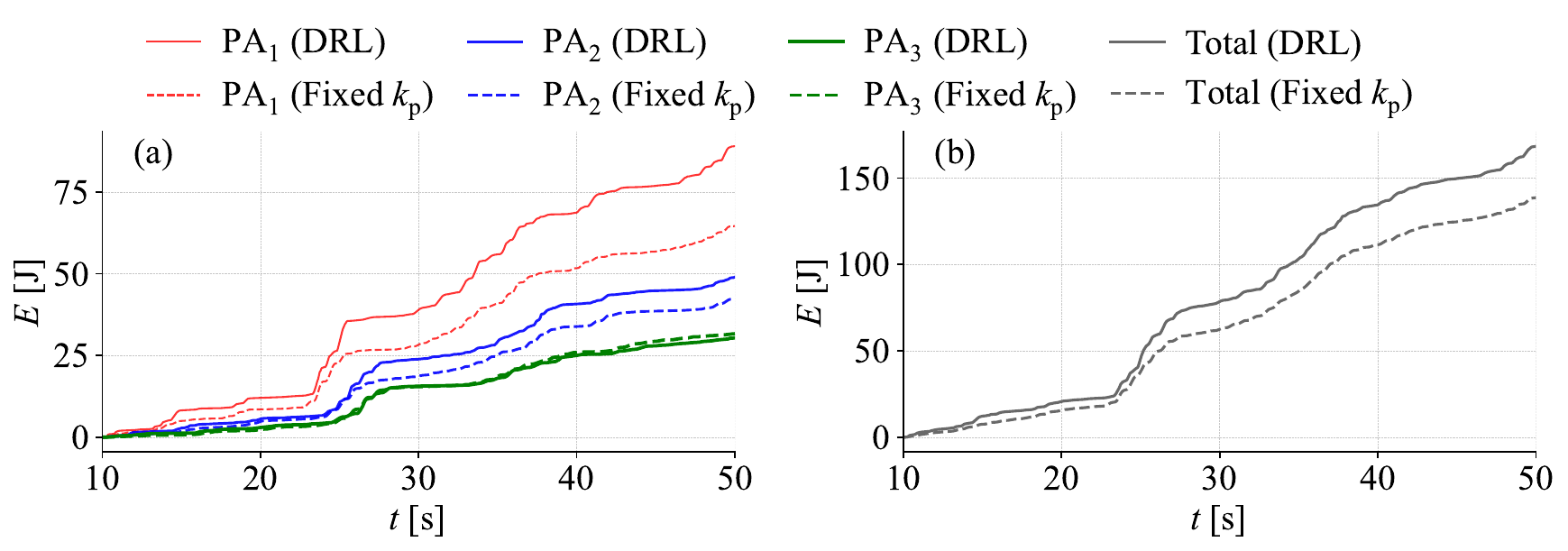}  
\caption{Time histories of the accumulated energy for three PAs under 3-D irregular wave conditions.}
\label{fig:irre_wec_energy}
\end{figure}

To quantitatively evaluate the improvement in energy-capture efficiency after training, the time histories of the instantaneous power output of each agent over the time interval from $t = 10$~s to $50$~s are presented in \Cref{fig:irre_wec_energy}. From the figure, the increase in energy is nonlinear due to the irregular distribution of wave energy; for instance, a rapid rise in power output is observed around $t \approx 25$~s, which corresponds to higher instantaneous surface elevation events in the irregular wave train. \Cref{tab:irre_energy_comparison} compares the total captured energy within 50~s before and after training. Compared with the constant $k_{\text{pto}}$ baseline, both PA$_1$ and PA$_2$ exhibit significant improvements in energy output, with increase ratios of 37.7\%, 15.7\%, and $-4.0$\% for PA$_1$ to PA$_3$, respectively. A slight decrease in PA$_3$ is observed, mainly due to the reward function being designed to maximise the overall array performance; given that the residual wave energy is already small when reaching PA$_3$, increasing its individual absorption contributes less to the global reward than improving the performance of PA$_1$ and PA$_2$. Overall, the total energy output of the entire PA array continues to increase, indicating that the DRL-trained agents can effectively cooperate to enhance the collective wave-energy absorption efficiency.

\begin{table}[htbp]
\centering
\caption{Comparison of the accumulated energy of three PAs over a 40~s simulation under 3-D irregular wave conditions.}
\label{tab:irre_energy_comparison}
{\renewcommand{\arraystretch}{1.2}
\begin{tabular}{ccccc}
\toprule
Agent & $E_{\text{DRL}}$ (J) & $E_0$ (J) & $\Delta E$ (J) & Improvement (\%) \\
\midrule
PA$_1$ & 89.03 & 64.64 & 24.39 & +37.7 \\
PA$_2$ & 49.00 & 42.34 & 6.66  & +15.7 \\
PA$_3$ & 30.38 & 31.66 & -1.28 & -4.0  \\
\midrule
Total  & 168.42 & 138.64 & 29.78 & +21.5 \\
\bottomrule
\end{tabular}}
\end{table}

\section{Concluding remarks and future perspectives}
\label{sec:conclusions}
In this study, a coupled platform integrating Smoothed Particle Hydrodynamics (SPH) with multi-agent deep reinforcement learning (MADRL) was developed and applied to the numerical simulation of point-absorber arrays. The framework dynamically adjusts the damping coefficient of the power take-off (PTO) system to enhance wave-energy harvesting performance. The SPH simulations were carried out using an extended version of the DualSPHysics solver, namely, DualSPHysics+, which includes additional numerical schemes for improved stability and accuracy. The reinforcement-learning component is implemented using the LibTorch framework, dynamically linked to the SPH solver to enable real-time exchange of hydrodynamic states and control actions. A centralised-training, decentralised-execution (CTDE) architecture is employed to achieve coordinated multi-agent control, addressing the convergence limitations observed in single-agent strategies under spatially varying wave conditions. The MASAC algorithm is employed to update the actor and critic neural networks in a stable manner. The entire system is implemented in CUDA C++, allowing GPU-based acceleration for both SPH computations and neural-network training, significantly improving computational efficiency and enabling the simulation of realistic three-dimensional engineering applications.

Two benchmark tests were conducted to assess the accuracy and stability of the SPH solver: wave propagation in an empty tank and the interaction between regular waves and a point absorber WEC. The results demonstrate that the model reproduces with accuracy the 2-D regular waves and the 3-D irregular wave conditions tested. In addition, the simulated heave velocity and displacement of the point absorber agree closely with the experimental data, illustrating the reliability of the model for wave–floating body interactions. 

The coupled SPH–MADRL framework was then applied to the adaptive adjustment of the PTO damping coefficients for point absorbers. In this setup, the SPH solver provides high-resolution flow-field data that serve as the local observations for each agent (PA). The MADRL model processes these observations through each agent’s policy neural network, which outputs incremental updates to the PTO damping coefficient. During training, the neural networks are updated following the SAC algorithm, and the resulting PTO coefficients are fed back into the SPH solver to perform the wave–floating body interaction simulation. 

The training results demonstrate that for single, multiple, and both two- and three-dimensional configurations, the cumulative reward increases steadily and ultimately converges, indicating that the learning process is both stable and reliable. The DRL-trained PAs develop an adaptive control policy whereby the PTO coefficient is increased during phases of high heave velocity, typically occurring near the passage wave crests and troughs and reduced during intermediate phases. This dynamic modulation enables the devices to attain larger heave velocities, displacements and PTO forces compared with the fixed-coefficient baseline, thereby enhancing instantaneous absorbed power. Quantitatively, the two-dimensional simulations show that the energy output of a single PA increases by 9.2\%, while arrays of two and three PAs achieve gains of 18.9\% and 23.8\%, respectively. In three-dimensional irregular wave conditions, the total energy output increases by approximately 21.5\%, demonstrating the model’s capability for large-scale, three-dimensional intelligent control and optimisation.

Beyond the global performance improvements, the results also reveal distinct spatial variations in energy distribution within the array. In the three-dimensional irregular wave array case, wave energy progressively attenuates along the propagation direction due to wave blocking and absorption by the upstream PAs. After optimisation through active control, the energy outputs of the individual PAs increase by 37.7\%, 15.7\%, and -4.0\% compared with the baseline case. Following the coordinated adjustment of the PTO coefficients across the array, a slight decrease in power output is observed for the downstream unit, as most of the incident wave energy has already been extracted by the upstream absorbers. This behaviour highlights the spatial attenuation of wave energy within the array and suggests that enhancing the energy-capture capability of upstream devices plays a more dominant role in improving overall array efficiency. Furthermore, these findings demonstrate that the multi-agent cooperative optimisation framework increases array-level power output by enabling adaptive energy allocation and coordinated adjustment of PTO coefficients across the array.
Note that this study presents the first demonstration of a fully integrated, GPU-accelerated SPH–MADRL framework capable of real-time two-way coupling and coordinated control of multi-body WEC arrays, addressing limitations of previous single-agent or 2-D CFD-based approaches.

Future work will extend the proposed framework to larger wave energy arrays involving a greater number of point absorbers, thereby enabling simulations that better represent realistic engineering applications. As the number of agents increases, the computational complexity of coordinated control and hydrodynamic interaction modelling grows rapidly. To address this issue, future work will focus on improving computational efficiency through multi-GPU parallelisation of both the SPH and DRL modules. In addition, future developments will aim to enhance the robustness of the control policies under sparse or noisy observations and delayed reward feedback, conditions that frequently arise in realistic ocean environments due to limited sensing and communication capabilities.

\appendix
\section{Computational efficiency analysis}
\label{app1}

This appendix presents the computational efficiency analysis for the 2-D case (consisting of three point absorbers) and the 3-D case introduced in \Cref{sec:adaptive_optimisation_pto_pa_arrays}. All simulations were performed on a micro-workstation; the corresponding hardware and software configurations are listed in \Cref{tab:workstation}. 

\begin{table}[htbp]
\centering
\caption{Workstation information for computational efficiency tests.}
\label{tab:workstation}
{\renewcommand{\arraystretch}{1.2}
\begin{tabular}{ll}
\toprule
Item & Specification \\
\midrule
CPU & Intel Xeon Gold 6133, 2.5\,GHz, 32 cores \\
GPU & NVIDIA RTX 4090, 24\,GB memory \\
RAM & 384\,GB DDR4 \\
Operating system & Ubuntu 20.04 LTS \\
Compiler & g++ 11.4, nvcc 12.8.93 \\
Parallel libraries & CUDA 12.8, OpenMP 4.5 \\
DRL library & LibTorch 2.8.0 \\
\bottomrule
\end{tabular}}
\end{table}

The initial particle spacing $dp$, total number of particles $N_{p}$, physical simulation time $t_{p}$ and runtime $t_{c}$ are summarised in \Cref{tab:simulation_summary}. It can be observed that the SPH computations still dominate the overall runtime, accounting for approximately 82.6\% and 87.6\% of the total cost in the 2-D and 3-D cases, respectively. It is noted that in the present model, a relatively small neural network that comprises only three fully connected layers ((128,128,64), as listed in \Cref{tab:masac_hyperparameters}) is sufficient to train the agents of this study. As the number of agents increases or larger neural networks are employed, the proportion of computational effort associated with the DRL component is expected to increase.

\begin{table}[htbp]
\centering
\caption{Summary of simulation parameters and computational performance for the 2-D and 3-D cases.}
\label{tab:simulation_summary}
{\renewcommand{\arraystretch}{1.2}
\begin{tabular}{lcccccc}
\toprule
\multirow{2}{*}{} & \multirow{2}{*}{$dp$ (m)} & \multirow{2}{*}{$N_p$} & \multirow{2}{*}{$t_p$ (s)} & \multicolumn{3}{c}{$t_c$ (h)} \\
\cmidrule(lr){5-7}
 &  &  &  & SPH & DRL & Total \\
\midrule
2-D regular wave & 0.01 & 151,230 & 1000 & 7.84 & 1.65 & 9.49 \\
3-D regular wave & 0.02 & 1,973,090 & 1000 & 32.85 & 4.65 & 37.51 \\
\bottomrule
\end{tabular}}
\end{table}

\bibliographystyle{elsarticle-harv} 
\bibliography{./bib/ref}

\end{document}